\documentclass[useAMS,usenatbib]{mn2e}
\usepackage{amsmath,fleqn,graphicx,amssymb}
\usepackage{multirow}
\renewcommand{\[}{\begin{equation}}
\renewcommand{\]}{\end{equation}}
\def\p{\partial}\def\i{{\rm i}}

{\newif\ifnotend
\notendtrue
\def\veclist{ABCDEFGHIJKLMNOPQRSTUVWXYZabcdefghijklmnopqrstuvwxyz.}
\def\top#1#2.{#1}
\def\tail#1#2.{#2.}
\loop\expandafter\xdef\csname v\expandafter\top\veclist\endcsname%
{{\noexpand\bf\expandafter\top\veclist}}
\edef\veclist{\expandafter\tail\veclist}
\if\veclist.\notendfalse\fi\ifnotend\repeat}
\def\d{{\rm d}}

\def\cR{{\cal R}}\def\cI{{\cal I}}\def\cK{{\cal K}}

\def\kpc{\,\mathrm{kpc}}

\def\kms{\,\mathrm{km\,s}^{-1}}

\def\msun{\,{\rm M}_\odot}

\def\pc{\,\mathrm{pc}}
\def\e{\mathrm{e}}

\def\fracj#1#2{{\textstyle{#1\over#2}}}

\def\Rg{R}\def\Omg{\Omega} 
\def\cL{{\cal L}}

\title[Swing amplification revisited]
{The shearing sheet and swing amplification revisited}

\author[James Binney]{
  James Binney$^1$\thanks{E-mail: binney@thphys.ox.ac.uk}\\  
  $^1$Rudolf Peierls Centre for Theoretical Physics, Clarendon Laboratory,
  Parks Road, Oxford, OX1 3PU, UK
}

\begin{document}
\maketitle

\begin{abstract}
The principal results of the classic analysis of the shearing sheet and swing
amplification by Julian \& Toomre (1966) are re-derived in a more accessible
way and then used to gain a better quantitative understanding of the dynamics
of stellar discs.  The axisymmetric limit of the shearing sheet is derived
and used to re-derive Kalnajs' 1965 dispersion relation and Toomre's 1964
stability criterion for axisymmetric disturbances.  Using the shearing sheet
to revisit Toomre's important 1969 paper on the group velocity implied by
Lin-Shu-Kalnajs dispersion relation, we discover that two rather than one
wavepackets emerges inside corotation: one each side of the inner Lindblad
resonance. Although LSK dispersion relation provides useful interpretations
of both wavepackets, the shearing sheet highlights the limitations of the LSK
approach to disc dynamics.  Disturbances by no means avoid an annulus around
corotation, as the LSK dispersion relation implies.  While disturbances of
the shearing sheet have a limited life in real space, they live on much
longer in velocity space, which Gaia allows us to probe extensively.  C++
code is provided to facilitate applications of winding spiral waves.
\end{abstract}


\section{Introduction} \label{sec:intro}

Spiral structure has fascinated astronomers since its discovery two
centuries ago. Although our quantitative grasp of this phenomenon remains
inadequate, we can now claim a good qualitative understanding of the
phenomenon, and we know that it plays a key role in the dynamical and chemical
evolution of galaxies like ours \citep[e.g.][]{Aumer2016a}.

The key ingredients of spiral structure are: the Lin-Shu-Kalnajs (LSK)
dispersion relation\footnote{This relation, first given by \cite{LinShu1966},
generalises the relation obtained for axisymmetric disturbances by
\cite{Kalnajs1965}.} for running density waves; the swing amplifier
\citep{GoldreichDLB,JT1966,Toomre1981}; and resonant absorption at Lindblad
resonances \citep{LLKa72,SellwoodC2014}. The picture is as follows. Noise
from any source will contain a packet of leading spiral waves. The LSK
dispersion relation is such that the packet will travel away from the nearest
Lindblad resonance towards corotation, unwinding as it goes \citep{ToomreGp}.
Eventually the tight-winding approximation on which the LSK relation is based
fails, and following \citet[][hereafter JT66]{JT1966} the packet's evolution
must be understood in the context of the shearing sheet. As the packet swings
from leading to trailing, it is amplified, and two stronger packets of
trailing waves propagate away from corotation towards the Lindblad
resonances. Should any part of these waves be reflected back towards
corotation as leading waves, repeated swing-amplification can permit
insignificant noise to grow into a manifest trailing spiral pattern and
ultimately a strong bar.

One way in which a trailing wave moving away from corotation can convert into
a leading wave that returns to corotation, is refraction around the galactic
centre \citep{Toomre1981}. However, this will not occur if a Lindblad
resonance is encountered before the centre is reached because the wave will
be resonantly absorbed there \citep{LLKa72} rather than proceeding to the
centre. Moreover, when the circular-speed curve is rather flat, any wave will
have a Lindblad resonance between corotation and the centre.

\cite{SellwoodC2014}, however, pointed out when a wave is resonantly absorbed
at its Lindblad resonance, a barrier is formed that is liable to reflect some
part of any subsequent wave as it approaches the barrier en route to its
Lindblad resonance. This is because absorption of the wave permanently
modifies the distribution function (DF) in a narrow zone around the
resonance, and it is generically the case that waves are partially reflected
when the impedance of the medium in which they are travelling changes in a
distance smaller than their wavelength.  Hence, if initially a disc's DF is a
smooth function of the actions, the first wavepacket reaches its Lindblad
resonance, where it is resonantly absorbed by Landau damping. As a result of
this absorption, the DF develops large gradients in the vicinity of the
Lindblad resonance and when a subsequent packet of swing-amplified waves
attempts to pass through this region of modified DF, it is partially
reflected back towards corotation, where it is again swing amplified.
Consequently this second wave packet modifies the DF more strongly at its
Lindblad resonances than did the first packet, and the scope for reflection
back to corotation for repeated amplification grows as the disc ages
\citep{SellwoodC2014}. Eventually the effectiveness of swing amplification is
such that the spiral structure becomes an order-unity phenomenon and
refashions the centre of the disc into a bar. The bar may later buckle into a
bulge/bar \citep{CombesSanders1981,Raha1991}.

The swing amplifier and resonant absorption are the stand-out pieces of
physics in this beautiful mechanism by which galaxies like ours evolve. The
first aim of this paper is to present an accessible account of the swing
amplifier, which \cite{GDII} shied away from attempting. They did so because
the derivation in JT66 is subtle and hard to follow. It relies on the theory
of characteristics, which will be unfamiliar to most students, and employs a
bewilderingly large number of symbols and changes of variables.  This paper
re-derives the key results of JT66 in a way that will be more accessible to
the average member of the community. With this pedagogic aim in view, more
intermediate steps in the algebra are given than are strictly necessary.

The second aim of the paper is to illustrate the value of the key result in
JT66 by plotting a variety of figures that help us to understand how stellar
discs work. Many of these figures appeared already in either JT66,
\cite{ToomreGp} or \cite{Toomre1981}, but with limited explanation of the
computational details. 

Section~\ref{sec:two} presents the shearing sheet from a Hamiltonian
perspective. Section \ref{sec:three} sets up the key equation by linearising
the collisionless Boltzmann equation (CBE). Section~\ref{sec:vspace} computes
the signature of a winding spiral in velocity space.
Section~\ref{sec:axisymm} derives the equation that governs axisymmetric
disturbances. This simpler equation admits modes and their dispersion
relation is derived and show to be identical with the axisymmetric limit of
the LSK dispersion relation.  Section~\ref{sec:four} uses the key equation to
investigate the response of the disc to an impulsive jolt.
Section~\ref{sec:mass} uses it to determine the increase in density around an
orbiting mass that arises because the disc is highly `polarisable', and
Section~\ref{sec:packets} uses it to study the dynamics of wavepackets.
Section~\ref{sec:discuss} discusses the insights we gain from the formalism
and its limitations. Section~\ref{sec:five} sums up. Appendix~\ref{app:A}
evaluates a required Gaussian integral, while Appendix~\ref{app:B} describes
a C++ class that computes the dynamics of winding waves.

\section{Shearing sheet}\label{sec:two}
\begin{figure}
\centerline{\includegraphics[width=.6\hsize]{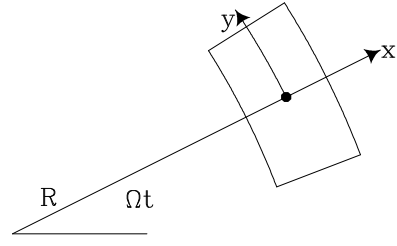}}
\caption{Schematic of the $(x,y)$ coordinate system for the shearing sheet.}
\label{fig:coords}
\end{figure}

We consider waves characterised by a large value of the azimuthal quantum
number $m$ and study a small near-Cartesian patch that moves around the disc
at the frequency $\Omega$ of a circular orbit at the mean radius $R$ of the
patch. We define a radial coordinate $x$ by taking a general radius to be
$\Rg+x$, and we define a complementary coordinate by $y\equiv\Rg\phi$,
where $\phi$ is the angle between the lines from the disc's centre to the
patch's centre and the point $(x,y)$ (Fig.~\ref{fig:coords}). In this
rotating frame, the Lagrangian is
\[
\cL=\fracj12\biggl[\dot x^2+(\Rg+x)^2\biggl(\Omg+{\dot y\over\Rg}\biggr)^2\biggr]-\Phi(\Rg+x)
\]
because in an inertial frame the angular velocity of the point at  $R+x$ is
$\Omega+\dot\phi=\Omega+\dot y/(\Rg+x)\simeq\Omega+\dot y/\Rg$. From $\cL$ we can read off 
the momenta:
\begin{align}
p_x&=\dot x\cr
p_y&=(\Rg+x)^2\left(\Omg+{\dot y\over\Rg}\right){1\over\Rg}\cr
&\simeq \Rg\Omg+2\Omg x+\dot y.
\end{align}
It follows that the Hamiltonian is
\begin{align}
H&=\fracj12\left( p_x^2+{p_y^2\over(1+x/\Rg)^2}\right)-\Omg\Rg p_y+\Phi.
\end{align}
 Since $H$ doesn't contain $t$ or $y$, we have two constants of motion: $H$ and $p_y$ or 
\[\label{eq:Delta}
\Delta_y\equiv p_y-\Rg\Omg=2\Omg
x+\dot y.
\]
 We will use $\Delta_y$ rather than $p_y$ because, unlike $p_y$, it is
first-order in the small quantities $x,y$.

The derivatives of $\Phi$ at the origin are
\begin{align}
{\p\Phi\over\p x}&=R\Omega^2\cr
{\p^2\Phi\over\p x^2}&=\Omg^2+2\Rg\Omg{\p\Omg\over\p\Rg}\cr
&=\Omg\left(\Omg-4
A\right),
\end{align}
 where $A$ is Oort's first constant (Table~\ref{tab:Oort}). Hence
\[
\Phi(\Rg+x)\simeq\Phi(\Rg)+\Rg\Omg^2 x+\fracj12\Omg(\Omg-4A)x^2,
\]
so
\begin{align}\label{eq:H}
H&\simeq\fracj12\left[p_x^2+p_y^2\left(1-2{x\over\Rg}+3{x^2\over\Rg^2}\right)\right]-\Rg\Omg
p_y+\Phi(\Rg)\cr
&\hskip3cm+\Rg\Omg^2x+\fracj12\Omg(\Omg-A)x^2\cr
&\simeq\fracj12\left[p_x^2+\Delta_y^2-\Rg^2\Omg^2\right]+\Phi(\Rg)-x\Omg\Delta_y+\fracj12\kappa^2x^2
\end{align}
where $\kappa^2\equiv4\Omg(\Omg-A)$.
From the way $x$ and $p_x$ appear in $H$ if follows that $x$ oscillates
harmonically  at the epicycle frequency $\kappa$ about
\[\label{eq:xbar}
\overline{x}\equiv2\Omg\Delta_y/\kappa^2.
\]

\begin{table}
\caption{Relations between frequencies. Fundamentally, there are just two
frequencies in the problem, $\Omega$ and $\d\Omega/\d\ln R$, but it proves
expedient from these to  define three mutually dependent
frequencies.}\label{tab:Oort} 
\centerline{
{\hsize=.6\hsize\vbox{
\hrule
\begin{align}
A&\equiv-\fracj12{\p\Omega\over\p\ln R}\qquad B\equiv A-\Omega\cr
\kappa^2&=4\Omega(\Omega-A)=-4\Omega B
\nonumber\end{align}
\hrule
}}}
\end{table}

\subsection{Orbits}

 Circular orbits are ones on which $x=\overline{x}$, so from equation
 (\ref{eq:Delta})
\begin{align}\label{eq:circ}
\dot y&=\Delta_y-2\Omg x=\left({\kappa^2\over2\Omg}-2\Omg\right)x\cr
&=-2Ax
\ \hbox{(circular orbit)}.
\end{align}
This relation describes how the sheet shears in the absence of random
velocities.

 Let $v_\phi\equiv2Ax+ \dot y$ be the azimuthal speed relative to the
local circular orbit. Then from equation (\ref{eq:xbar}) and the definition
(\ref{eq:Delta}) of $\Delta_y$ we have
\[
v_\phi=2(A-\Omg)x+{\kappa^2\overline{x}\over2\Omg}=2B(x-\overline{x}),
\]
 where $B$ is Oort's second constant (Table~\ref{tab:Oort}).
Hence $v_\phi$ tells us how far a star is from its guiding centre.
Given that $x$ oscillates harmonically, the coordinates of any star can be
written
\begin{align}\label{eq:genorb}
x&=\overline{x}+X\cos\theta_r\quad\Rightarrow\quad p_x=-\kappa X\sin\theta_r
\cr
v_\phi&=2B(x-\overline{x})=2BX\cos\theta_r.
\end{align}
where $\theta_r=\kappa t+\hbox{constant}$. Similarly,
\begin{align}\label{eq:genorb2}
\dot y&=\Delta_y -2\Omg x=\Delta_y -2\Omg(\overline{x}+X\cos\theta_r)\cr
\Rightarrow\ 
y(t')&=y_0+\Delta_y\left(1-{4\Omega^2\over\kappa^2}\right)t'-2{\Omg\over\kappa}X\sin\theta_r\cr
&=y_0+\Delta_y{A\over B}t'-2{\Omg\over\kappa}X\sin\theta_r,
\end{align}
where $y_0$ is a constant of integration.

 When we eliminate $x$ from $H$ in favour of $v_\phi$ we get
\begin{align}
H&\simeq\fracj12\left[p_x^2+\Delta_y^2-\Rg^2\Omg^2\right]+\Phi(\Rg)
+\fracj12\kappa^2\left[(x-\overline{x})^2-\overline{x}^2\right]\cr
&=\fracj12\left[p_x^2+\Delta_y^2\left(1-{4\Omg^2\over\kappa^2}\right)-\Rg^2\Omg^2\right]+\Phi(\Rg)
+{\kappa^2\over 8B^2}v_\phi^2\cr
&=\fracj12\left[p_x^2+\Delta_y^2{A\over A-\Omg}-\Rg^2\Omg^2\right]+\Phi(\Rg)
+{\Omg\over -2B}v_\phi^2\cr
&=H_x(p_x,v_\phi)+H_y(\Delta_y),
\end{align}
where
\[\label{eq:Hval}
H_x\equiv\fracj12\left(p_x^2+{\Omega\over-B}v_\phi^2\right)=\fracj12\kappa^2X^2.
\]
Since $B<0$ and $p_x$ is the radial component of velocity, in velocity space
lines of constant $H_x$ form ellipses.  From equation (\ref{eq:H}) we obtain
an alternative expression for $H_x$:
\[\label{eq:HOx}
H_x=\fracj12\left[p_x^2+\kappa^2(x-\overline{x})^2\right].
\]

\begin{figure}
\centerline{\includegraphics[width=.8\hsize]{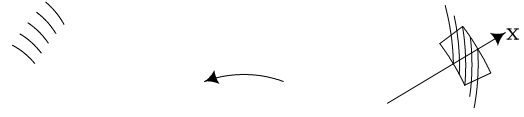}}
\caption{A washboard of leading waves is sheared into radially directed
crests as the coordinate patch is carried around the disc.}
\label{fig:shear2}
\end{figure}

\subsection{Characteristic scales}

By Jeans' theorem a suitable equilibrium DF is
\[\label{eq:defs_f0}
f_0=F\e^{-H_x/\sigma^2},
\]
where $F$ and $\sigma$ are free constants. By equation (\ref{eq:Hval}) this
DF generates a biaxial Maxwellian velocity distribution with dispersion
$\sigma$ in $p_x$ and $\sqrt{-B/\Omega}\sigma$ in $v_\phi$.  The
surface density it generates  is
\[\label{eq:surfdens}
\Sigma_0=\int\d p_x\d
v_\phi\,f_0={4\pi\sigma^2BF\over\kappa}.
\]

It  is convenient to specify the constants $F$ and $\sigma$ through the
values of two functions of them. \cite{To64} showed that even a stellar disc in which all
stars are on perfectly circular orbits (so $\sigma=0$) is stable to
axisymmetric perturbations with wavenumbers smaller than  the critical wavenumber 
\[\label{eq:defs_kcrit}
k_{\rm crit}\equiv{2\pi\over\lambda_{\rm crit}}={\kappa^2\over2\pi
G\Sigma_0}.
\]
We use $k_{\rm crit}$ as proxy for the surface density $\Sigma_0$ or,
equivalently, the
normalisation of the DF, $F$. Our proxy for $\sigma$ is
\[\label{eq:defs_Q}
Q\equiv {\kappa\sigma\over3.36G\Sigma_0},
\]
which is the ratio of the actual velocity dispersion $\sigma$ to the minimum value
that ensures that the disc is stable to axisymmetric disturbances of any
wavenumber \citep{To64}.

If the disc is disturbed by a potential $\Phi_1\propto\cos(k_yy)$, a star
that is on a circular orbit at $x$ experiences the disturbance at frequency
\[
\omega=k_y\dot y=-2Axk_y.
\]
At the locations $\pm x_{\rm L}$ of the Lindblad resonances, the absolute
value of this frequency coincides with the star's radial frequency $\kappa$.
Clearly the Lindblad resonances of this disturbance lie at
\[\label{eq:defs_xL}
x_{\rm L}=\pm{\kappa\over2Ak_y}.
\]

If we adopt plausible values $\Sigma_0=49\msun\pc^{-2}$ and
$\kappa=37\kms\kpc^{-1}$ for the solar neighbourhood \citep{GDII}, then
$\lambda_{\rm crit}\simeq6\kpc$ and $k_{\rm crit}\sim1\kpc^{-1}$.

\subsection{Shearing pattern}

We assume that our coordinate patch is centred on the corotation radius of a
spiral wave as sketched on the right of Fig.~\ref{fig:shear2}. Hence the
pattern shears with particles on circular orbits. That is, if the perturbed
surface density has the form
\[
\Sigma_1(\vx,t)=\widetilde\Sigma_1(t)\e^{\i\vk\cdot\vx},
\] 
 the phase $\vk\cdot\vx$ is constant at the location of every particle that
moves on a circular orbit. For this to be so, $\vk$ has to be a function of
time. We determine that function by equating phases at a particle's locations at two
different times:
\begin{align}
&k_x(t_\i)x+k_yy_0=k_x(t)x+k_yy(t)\cr
\Rightarrow\quad&
k_x(t)=k_x(t_\i)+2k_yA(t-t_\i),
\end{align}
 where we have used equation (\ref{eq:circ}).
Henceforth we take the origin of time to be the instant 
\[
t_c\equiv-{k_x(t_\i)\over 2k_yA},
\] 
at which $k_x=0$. At times $t<0$, $k_x<0$ and the spiral is leading, while at
$t>0$ it is trailing. In general
 \[\label{eq:kx}
k_x(t)=2Ak_yt,
\]
 and 
\[\label{eq:modk}
|k|=k_y\sqrt{1+4A^2t^2}
\]
 goes through a minimum as $k_x$ passes
through zero.

So long as $|k|\ll R$, the gravitational potential generated by a
sinusoidal density variation in a razor-thin disc is \citep[e.g.][]{GDII}
\[\label{eq:Poisson}
\Phi_1(\vx,t)=P(t)\e^{\i\vk\cdot\vx} =-{2\pi G\Sigma(\vx,t)\over|k|}.
\]
The strength of this potential peaks as $k_x$ passes through zero.

Using our solution (\ref{eq:genorb}) and (\ref{eq:genorb2}) for a general orbit we now have that
\begin{align}
k_xx+k_yy&=2k_yAt(\overline{x}+X\cos\theta_r)\cr
&\qquad+k_y\left(y_0+\Delta_y{A\over
B}t-2{\Omg\over\kappa}X\sin\theta_r\right)\cr
&=k_y\left[y_0+2X\left(At\cos\theta_r-{\Omg\over\kappa}\sin\theta_r\right)
\right].
\end{align}
If we define the phase 
\[\label{eq:psi}
\psi(t)\equiv2k_yX(At\cos\theta_r-(\Omg/\kappa)\sin\theta_r).
\]
this simplifies to
\[\label{eq:Usepsi}
\vk\cdot\vx=k_yy_0+\psi(t).
\]
It follows that if $t$ and $t'$ are two instants along a given
orbit, then
\[\label{eq:diffpsi}
\vk\cdot\vx|_{t'}=\vk\cdot\vx|_t+\psi(t')-\psi(t).
\]

\section{Linearizing the CBE}\label{sec:three}

With $f=f_0(H_0)+f_1$ and $H=H_0+\Phi_1$, the CBE
\[
{\p f\over\p t}+[f,H]=0,
\]
 where $[,]$ is a Poisson bracket, linearises to
\[
{\d f_1\over\d t}\equiv{\p f_1\over\p t}+[f_1,H_0]=[\Phi_1,f_0].
\]
 On integration we have
\[\label{eq:along}
f_1=\int_{t_\i}^t\d t'\,{\p\Phi_1\over\p\vx}\cdot{\p f_0\over\p\vp},
\]
 where the integral is along unperturbed orbits and $f_1$ has been assumed to
vanish for $t'<t_\i$.

The unperturbed  DF (\ref{eq:defs_f0}) is a function of not only $p_x,x$ but
through $\overline{x}$ or $v_\phi$ also
of the momentum $\Delta_y$. Using equation (\ref{eq:HOx}) for $H_x$
with
$\Phi_1(\vx)\propto\e^{\i\vk\cdot\vx}$, we have 
\[\label{eq:integrand}
{\p\Phi_1\over\p\vx}\cdot{\p f_0\over\p\vp}=
-\i
F\Phi_1{\e^{-H_x/\sigma^2}\over\sigma^2}[k_xp_x-k_y2\Omg(x-\overline{x})].
\] 

Since we have assumed that $f_1$ vanishes at $t_\i$, it
will vanish at all subsequent time unless something contributes to $\Phi_1$
in addition to the surface density associated with $f_1$. Hence,
we take
\begin{align}\label{eq:kick}
&\Phi_1(t')=-{2\pi G\over|k|}
\left[\widetilde\Sigma_\e(t')+\widetilde\Sigma_1(t')\right]\e^{\i\vk\cdot\vx|_{t'}}\cr
&=-{2\pi G\over|k|}\!
\left[\widetilde\Sigma_\e(t')+\widetilde\Sigma_1(t')\right]\e^{\i[\vk\cdot\vx|_t+\psi(t')-\psi(t)]},
\end{align}
 where the second equality uses equation (\ref{eq:diffpsi}),
$\widetilde\Sigma_\e(t')\e^{\i\vk\cdot\vx|_{t'}}$ is a fictitious surface density that provides a
gravitational driving
force, and
$\widetilde\Sigma_1$ is defined by
\[\label{eq:defwSigma}
\Sigma_1(\vx,t)=\int\d^2\vv\,f_1=\widetilde\Sigma_1(t)\e^{\i\vk\cdot\vx}.
\]

In view of equations (\ref{eq:along}) to (\ref{eq:kick}), with
equations (\ref{eq:genorb}) and (\ref{eq:modk})  we have
\begin{align}\label{eq:foneSigma}
f_1(t)&=2\pi GF\i{\e^{-H_x/\sigma^2}\over\sigma^2}\e^{\i\vk\cdot\vx}\int_{t_\i}^t\d
t'\,\bigl(\widetilde\Sigma_\e+\widetilde\Sigma_1\bigr)\cr
&\times
\e^{\i[\psi(t')-\psi(t)]}{-2At'\kappa X\sin\theta_r-2\Omg
X\cos\theta_r\over\sqrt{1+4A^2t^{\prime2}}}.
\end{align}

Next we have to compute $\widetilde\Sigma_1$ by integrating  $f_1$ over $\vv$,
but before integrating we eliminate
$X,\theta_r$ in favour velocity components $p_x,v_\phi$ at $t'=0$. Using $
\theta_r=\theta_0+\kappa t'$ we write
\begin{align} 
\kappa X\sin\theta_r&=\kappa X\left(\sin\theta_0\cos\kappa
t'+\cos\theta_0\sin\kappa t'\right)\cr
&=-U_x\cos\kappa t'+U_y\sin\kappa t',
\end{align}
where
\begin{align}\label{eq:U_pv}
 U_x&\equiv-\kappa X\sin\theta_0=p_x\cr
U_y&\equiv\kappa
X\cos\theta_0=(\kappa/2B)v_\phi.
\end{align}
Similarly
\[
\kappa X\cos\theta_r=U_y\cos\kappa t'+U_x\sin\kappa t'.
\]
The Jacobian of the (time-dependent) transformation in velocity is simply
\[\label{eq:Jacob}
{\p(p_x,v_\phi)\over\p(U_x,U_y)}={2B\over\kappa}.
\]
With $\vU$ replacing $(p_x,v_\phi)$ the phase factor of the integrand in equation
(\ref{eq:foneSigma}) becomes
\begin{align}\label{eq:psi_diff}
\psi(t')-\psi(t)&=2(k_y/\kappa)\biggl[U_x\Bigl\{A(t'S'-tS)+{\Omega\over\kappa}(C'-C)\Bigr\}\cr
&\ +U_y\Bigl\{A(t'C'-tC)-{\Omg\over\kappa}(S'-S)\Bigr\}\biggr],
\end{align}
where $C(t)\equiv\cos\kappa t$, $S(t)\equiv\sin\kappa t$ and $C'$ and $S'$
are defined the same way but with $t'$ replacing $t$.

Using equation (\ref{eq:surfdens}) to eliminate $F$ from equation (\ref{eq:foneSigma}), we
see that the
coefficient of $\e^{\i\vk\cdot\vx}$ in the expression for the
perturbed surface density  can be written
\begin{align} \label{eq:Sigone}
\widetilde\Sigma_1(t)&=\int\d p_x\d v_\phi\,\e^{-\i\vk\cdot\vx}\,f_1\cr
&=\int_{t_\i}^t\kappa\,\d
t'\,K(t,t')\left[\widetilde\Sigma_\e(t')+\widetilde\Sigma_1(t')\right],
\end{align}
where
\[\label{eq:defK}
K(t,t')\equiv{I(t,t')\over\sqrt{1+4A^2t^{\prime2}}}
\]
is a dimensionless kernel that involves the Gaussian integral
\begin{align}
I(t&,t')\equiv-2\i{G\Sigma_0\over\kappa\sigma^4}\int\d^2\vU\,\e^{-U^2/2\sigma^2}
\e^{\i[\psi(t')-\psi(t)]}\cr
&\times
\left[At'(-U_xC'+U_yS')+(\Omega/\kappa)(U_yC'+U_xS')\right].
\end{align}
The integral is of the form
\[
\int\d^2\vU\,\vU\cdot\vc\,\e^{-a^2U^2+2\i\vb\cdot\vU}=\i\pi{\vc\cdot\vb\over
a^4}\,\e^{-b^2/a^2},
\]
where
\begin{align}\label{eq:def_abc}
a^2&={1\over2\sigma^2}\cr
b_x&={k_y\over\kappa}\bigl[A(t'S'-tS)+(\Omg/\kappa)(C'-C)\bigr]\cr
b_y&={k_y\over\kappa}\bigl[A(t'C'-tC)-(\Omg/\kappa)(S'-S)\bigr]\cr
c_x&=-At' C'+{\Omg\over\kappa} S'\qquad
c_y=At' S'+{\Omg\over\kappa} C'.
\end{align}
Hence
\[
I(t,t')=8\pi{G\Sigma_0\over\kappa}\,\vc\cdot\vb\,\e^{-b^2/a^2}.
\]
In terms of the critical wavenumber (\ref{eq:defs_kcrit}),
$I$ simplifies to
\[\label{eq:Ifromb}
I(t,t')={4\kappa\over k_{\rm crit}}\,\vc\cdot\vb\,\e^{-b^2/a^2}.
\]
Now we replace $\vb$ by the dimensionless vector
\[
\widehat\vb\equiv{\kappa\over k_{\rm crit}}\vb
\]
and note that
\[\label{eq:introQ}
{b \over a }=\surd2\sigma\,{k_{\rm crit} \over\kappa}\widehat b={3.36\surd2\over2\pi}Q\widehat b,
\]
where we have introduced $Q$ (eqn.~\ref{eq:defs_Q}).
With equations (\ref{eq:Ifromb}) and (\ref{eq:introQ}), 
the kernel (\ref{eq:defK}) can be written
\[\label{eq:finalK}
K(t,t')=4\vc\cdot\widehat\vb\,{\exp\bigl(-0.572Q^2\widehat b^2\bigr)\over\sqrt{1+4A^2t^{\prime2}}}.
\]
The vectors $\widehat\vb$ and $\vc$ are functions of the dimensionless numbers $At$, $At'$,
$\kappa t$, $\kappa t'$
and $\Omega/\kappa$, while $\widehat\vb$ has an additional (linear) dependence on
$k_y/k_{\rm crit}$. The argument $(Q\widehat b)^2$ of the exponential is
proportional to $\sigma^2$ and has no dependence on $\Sigma_0$.

In the following we shall refer to equation (\ref{eq:Sigone}) as the JT
equation and $K$ as the JT kernel. Although JT66 normalised their kernel
differently, equation (\ref{eq:Sigone}) is equivalent to their equation.

\subsection{Impact on velocity space}\label{sec:vspace}

Once we have computed the evolution of $\widetilde\Sigma_1$ from the JT
equation, we can use equation (\ref{eq:foneSigma}) to compute  the evolution
of $f_1$. Since we can directly observe the phase space of our Galaxy,
$f_1(\vx,\vv,t)$ contains observationally significant information that was erased when
we integrated over $\vU$ to compute $\widetilde\Sigma_1$.

Equation (\ref{eq:foneSigma}) is explicit about the $\vx$ dependence of
$f_1$, but its $\vv$ dependence is buried in the variables $X$ and
$\theta_r$. Above we expressed $X$ and $\theta_r$ as functions of the velocity at
the special moment when $k_x=0$. Now we want to express them as functions of
the velocity at the current time $t$. We do this by now writing
$\theta_r=\theta_0+\kappa(t'-t)$, where $\theta_0$ is a star's current
radial phase. With this new definition of $\theta_0$, 
\begin{align}
\kappa X\sin\theta_r&=-U_x\cos[\kappa(t'-t)]+U_y\sin[(\kappa(t'-t)]\cr
\kappa X\cos\theta_r&=\ U_y\cos[\kappa(t'-t)]+U_x\sin[\kappa(t'-t)],
\end{align}
where $\vU$ is related to the current velocity components by equation
(\ref{eq:U_pv}). Equation (\ref{eq:psi_diff}) becomes
\begin{align}\label{eq:psi_diff2}
\psi(t')-\psi(t)&=2(k_y/\kappa)\biggl[U_x\Bigl\{A(t'-t)S'+{\Omega\over\kappa}(C'-1)\Bigr\}\cr
&\ +U_y\Bigl\{A(t'-t)C'-{\Omg\over\kappa}S'\Bigr\}\biggr]
\end{align}
where $C'=\cos[\kappa(t'-t)]$ and $S'=\sin[\kappa(t'-t)]$. Completing the
transition from $X,\theta_r$ to $\vU$, we find
\begin{align}\label{eq:foneSigma2}
f_1(t)&=-\i{\kappa\e^{-H_x/\sigma^2}\over3.36QB}\e^{\i\vk\cdot\vx}\int_{t_\i}^t\kappa\d
t'\,\bigl(\widetilde\Sigma_\e+\widetilde\Sigma_1\bigr)\e^{\i[\psi(t')-\psi(t)]}\cr
&\times
{A(t'-t)(-U_xC'+U_yS')+(\Omega/\kappa)(U_yC'+U_xS')
\over\sigma^3\sqrt{1+4A^2t^{\prime2}}}.
\end{align}

\subsection{Axisymmetric limit}\label{sec:axisymm}

We now recover the form of the JT equation that describes the response of a
disc to axisymmetric disturbances. That is, we take the limit $k_y\to0$
of the JT equation. Since $\widehat\vb\propto k_y$ and both the prefactor and the
exponent in equation (\ref{eq:finalK}) vanish with $k_y$, superficial
analysis yields the conclusion that in this limit $K\to0$, making the JT
equation useless. But we have to bear in mind that $k_x=2Atk_y$ and we
require $k_x\ne0$, so we must let $|t|\to\infty$ as we let $k_y\to0$ such
that the product $tk_y$ is constant. By allowing $t$ to diverge we are
recognising that the instant when $k_x=0$ and the wave crests run radially
lies in the remote past (or future if $k_x<0$).

A glance at  equations (\ref{eq:def_abc}) shows that in this limit $k_y\to0$
but $k_yt=\hbox{constant}$, $\vb$ and $\vc$ simplify such that
\begin{align}\label{eq:lim}
\widehat\vb&\to{1\over 2k_{\rm crit}}\bigl(k_x'S'-k_xS,k_x'C'-k_xC\bigr)\cr
{4\vc\over\sqrt{1+A^2t'^2}}&\to2(-C',S'),
\end{align}
where $k_x'\equiv k_x(t')$, etc. Now as $k_y\to0$ $\d k_x/\d t\to0$, so over
any finite interval $t-t'$ there will be negligible change in $k_x$.
Moreover, while we are obliged to let $|t,t'|\to\infty$, it suffices to
consider forces that acted a finite time in the past. That is, we need to
evaluate $K(t,t')$ for finite $|t-t'|$, so we can neglect the difference
between $k_x$ and $k_x'$. Simplifying the right sides of equations
(\ref{eq:lim}) thus and using the trigonometrical identities
$CC'+SS'=\cos[\kappa(t-t')]$ and $SC'-CS'=\sin[\kappa(t-t')]$, we find that
in the axisymmetric limit $k_y\to0$, the JT kernel reduces to (JT66)
\begin{align}\label{eq:axisymm}
\cK(t-t')&\equiv \lim_{k_y\to0}K(t,t')\cr
&={k_x\over k_{\rm crit}}\sin[\kappa(t-t')]\e^{(\cos[\kappa(t-t')]-1)\chi},
\end{align}
where
\[
\chi\equiv{Q^2k_x^2\over3.497 k_{\rm
crit}^2}=\Bigl({k_x\sigma\over\kappa}\Bigr)^2.
\]
The JT equation now reads
\[
\widetilde\Sigma_1(t)=\int_{-\infty}^t\kappa\,\d
t'\,\cK(t-t')\left[\widetilde\Sigma_\e(t')+\widetilde\Sigma_1(t')\right].
\]
We replace $t'$ by $t''\equiv t-t'$. As $t'$ goes from $-\infty$ to $t$,
$t''$ goes from $\infty$ to 0. Hence the equation becomes
\[
\widetilde\Sigma_1(t)=\int_0^{\infty}\kappa\,\d
t''\,\cK(t'')\left[\widetilde\Sigma_\e(t-t'')+\widetilde\Sigma_1(t-t'')\right].
\]
The right side has the form of a Laplace convolution of $\cK$ with the sum of
surface densities, so the Laplace transforms $\overline{\Sigma}_1(p)$, etc
are related by
\[
\overline{\Sigma}_1(p)={\kappa\overline{\cK}(p)\overline{\Sigma}_\e(p)
\over1-\kappa\overline{\cK}(p)}.
\]
Values of $p$ for which $\kappa\overline{\cK}=1$ are of particular interest: for
these frequencies the JT equation has a non-zero solution in the absence of
a stimulating density $\Sigma_\e$. That is, these are the frequencies of the
sheet's axisymmetric normal modes.

Using the identity \citep[][eqn.~9.6.34]{AbramStegun}
\[
\e^{z\cos\theta}=\sum_{n=-\infty}^\infty I_n(z)\cos(n\theta),
\]
it is easy to show that the Laplace transform of $\cK(\tau)$ is
\begin{align}\label{eq:enough}
\overline{\cK}(p)=&{k_x\e^{-\chi}\over 2k_{\rm crit}}
\sum_{n=-\infty}^\infty I_n(\chi)\biggl({(n+1)\kappa\over
p^2+(n+1)^2\kappa^2}\cr
&\qquad\qquad-{(n-1)\kappa\over p^2+(n-1)^2\kappa^2}\biggr).
\end{align}
When $n=-|n|$ the second term in the big bracket above takes the value that
the first term takes when $n=|n|$, so we can drop the second term and double
the remaining sum:
\[
\overline{\cK}(p)={k_x\e^{-\chi}\over k_{\rm crit}}
\sum_{n=-\infty}^\infty I_n(\chi){n+1\kappa\over p^2+(n+1)^2\kappa^2}.
\]
We relate this to the standard LSK dispersion relation by defining the
dimensionless frequency $s\equiv \i p/\kappa$. In terms of $s$ the dispersion
relation $\kappa\overline{\cK}(p)=1$ can be written
\[\label{eq:JTdisp}
0=1-{k_x\over k_{\rm crit}}\e^{-\chi}
\sum_{n=-\infty}^\infty I_n(\chi){(n+1)\over (n+1)^2-s^2}.
\]
When the LSK dispersion relation is written in analogous form
\citep[][eqn.~K.25]{GDII}, and  divided by $1-s^2$, one obtains
\[\label{eq:LSKdisp}
0=1-{|k|\over k_{\rm crit}}\e^{-\chi}
\sum_{n=-\infty}^\infty {I_n(\chi)\over\chi}{n^2\over
n^2-s^2}.
\]
 The recurrence relation
\[
I_{n-1}(\chi)-I_{n+1}(\chi)=2n{I_n(\chi)\over\chi}
\] 
allows us put the LSK relation into the form
\[
0=1-{|k|\over 2k_{\rm crit}}\e^{-\chi}
\sum_{n=-\infty}^\infty [I_{n-1}(\chi)-I_{n+1}(\chi)] {n\over
n^2-s^2}.
\]
Gathering together the coefficients of $I_n$ makes the sum precisely the sum
that appears in equation (\ref{eq:enough}), so the same manoeuvre to simplify
the coefficient of $I_n$ establishes that the LSK dispersion relation can be
written in the form (\ref{eq:JTdisp}) that was obtained from the JT kernel.

Passing to the axisymmetric limit removes all meaning to the concepts
`pattern speed' and `corotation'. Before we take this step, $x=0$ is singled
out as the location at which stars move at a fixed phase in the perturbation.
After going axisymmetric,  every value of $x$ is equivalent. Hence the
axisymmetric modes of sheets are standing sinusoidal waves of unlimited
extent in $x$. 

When the right-hand side of equation (\ref{eq:JTdisp}) is plotted as a
function of $k_x/k_{\rm crit}$ at a fixed values of $\sigma$ and $s^2$, one
obtains a curve that has a minimum at $k/k_{\rm crit}\sim2$. The smaller the
value of $\sigma$ or the larger the value of $s^2$, the lower the curve
reaches. If the curve crosses the line $y=0$, the system has a mode of
frequency $s$ with the
associated value of $k_x/k_{\rm crit}$. The system is stable if the curve
crosses $y=0$ only for $s^2>0$, and Toomre's stability criterion $Q>1$
emerges by finding the value of $\sigma$ at which the
curve just touches $y=0$ for infinitesimal positive  $s^2$.
The sheet's modes  oscillate indefinitely if stable ($s^2>0$) or they grow
exponentially when unstable: there are no overstable modes.

\section{Impulsive excitation of the disc}\label{sec:four}

\begin{figure}
\centerline{\includegraphics[width=\hsize]{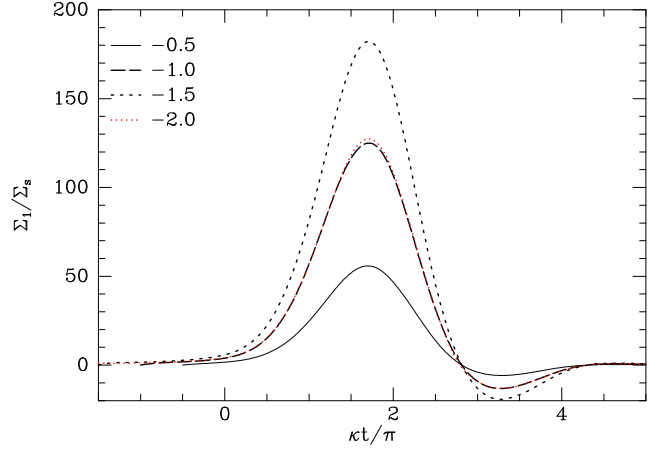}} \caption{Evolution
of waves with $k_y=k_{\rm crit}$ in a Mestel disc with $Q=1$ when stimulated
at $\kappa t_\i=-\pi/2,-\pi,-3\pi/2,-2\pi$ by the surface density
(\ref{eq:impulseS}). The legend at top left gives the relevant values of
$\kappa t_\i/\pi$ (cf.\ Fig.~4 of JT66).} \label{fig:jt_first}
\end{figure}

For our first application of the JT equation we follow JT66 by exciting the
disc impulsively. We take 
\[\label{eq:impulseS}
\widetilde\Sigma_\e(t)={\Sigma_{\rm s}\over\kappa}\delta(t-t_\i),
\]
 where $\Sigma_{\rm s}$ sets the magnitude of the impulse and the
denominator $\kappa$ ensures dimensional soundness.
With this choice, the JT equation yields
\[\label{eq:SigQ}
\widetilde\Sigma_1(t)=K(t,t_\i)\Sigma_{\rm s}+
\int_{t_\i}^t\kappa\d t'\,K(t,t')\widetilde\Sigma_1(t').
\]
Equation (\ref{eq:SigQ}) is straightforwardly solved by establishing a grid
of about a thousand equally spaced times in the interval $(t_\i,5\pi/\kappa)$,
approximating the integral over $t'$ by the trapezium rule on this grid, and
then for each possible end time $t$ solving the resulting algebraic equation
for $\widetilde\Sigma_1(t)$.

\begin{figure}
\centerline{\includegraphics[width=\hsize]{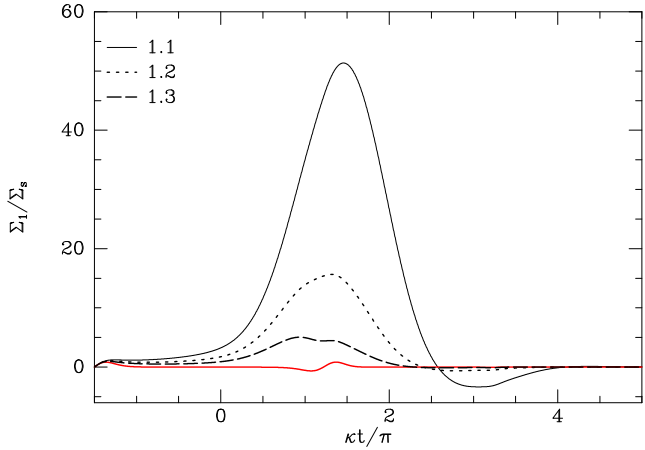}}
\caption{As Fig.~\ref{fig:jt_first} but for $\kappa t_\i=-3\pi/2$ and three values
of $Q$, namely $1.1,\,1.2$ and $1.3$ as indicated at top left. The red curve
shows the response when the disc's self gravity is neglected.}
\label{fig:jt_second}
\end{figure}

\begin{figure}
\centerline{\includegraphics[width=\hsize]{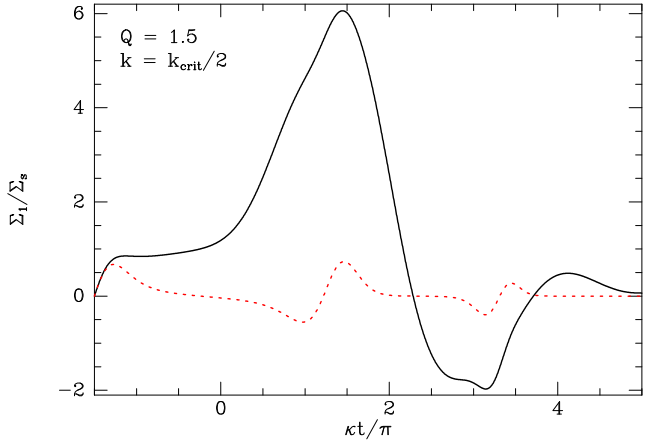}}
\caption{As Fig.~\ref{fig:jt_second} but for
of $Q=2$ and $k-k_{\rm crit}/2$. For these values swing amplification is less
effective and the complex structures of both the responses with (black) and
without (red) self-gravity can be examined on a common scale.}
\label{fig:jt_secondb}
\end{figure}

Fig.~\ref{fig:jt_first} shows solutions obtained in this way for four values
of $t_\i$ in a disc with a perfectly flat rotation curve, $Q=1$ and
$k_y=k_{\rm crit}$.  Since the applied stimulus
perturbs velocities but not positions, time is required for an overdensity to
develop that can be amplified as $k_x$ passes through zero at $t=0$. Hence
earlier excitation produces a bigger response up to times later than
$t_\i\simeq-3\pi/2\kappa$.  However, moving $t_\i$ further back decreases the
peak in $\widetilde\Sigma(t)$ because the disturbance then has time to phase
mix away prior to $t=0$. As a reflection of this tradeoff between phase
mixing and the need for time for fluctuations in density to emerge, in
Fig.~\ref{fig:jt_first} the red dotted curve for $\kappa t_\i=-2\pi$ almost
exactly coincides with the black, long-dashed curve for $\kappa t_\i=-\pi$.

The black curves in Fig.~\ref{fig:jt_second} show, for a fixed value of
$k_y$, the evolution of
$\widetilde\Sigma_1$ for $t_\i=-3\pi/2\kappa$ and three values of $Q$, namely
$1.1,\,1.2$ and $1.3$. As expected, as self-gravity becomes less important,
the amplitude of the response declines steeply. Nonetheless, self-gravity
greatly enhances the disc's response even for the largest plotted value of
$Q$. This fact is established by the red curve in Fig.~\ref{fig:jt_second},
which plots $K(t,t_\i)$ for $Q=1.3$, which is what the response would be
in a disc of test particles. 

The full black and dashed red curves in Fig.~\ref{fig:jt_secondb} show,
respectively, the responses with and without self-gravity for the case
$Q=1.5$ and $k=k_{\rm crit}/2$. In this warmer sheet and at this longer
wavelength,  swing amplification is less powerful and the complex responses
can be studied on a common scale. Notwithstanding the comparative weakness of
self-gravity in this case, it dramatically modifies the structure
of the response, not just its amplitude. In the absence of self-gravity, the response
consists of a quickly achieved maximum, followed by  series of blips, each of
which comprises a minimum quickly followed by a maximum. The first blip is
centred on $\kappa t=5\pi/4$ and the second on and $10\pi/3$. With self-gravity, the response
comprises a rapid rise to a plateau at a value slightly larger than the peak
response achieved without self-gravity, followed by a steep rise to a high
peak at $\kappa t=5\pi/4$ that commences just after $k_x$ passes through zero. This peak is then
rapidly followed by an almost double-bottomed minimum. The self-gravitating
response begins to  rise out this minimum at essentially the same moment that
the response without self-gravity starts to rise from its second minimum. The
self-gravitating response has a second maximum at $\kappa t\simeq4\pi$,
when without self-gravity the response is essentially zero.

The key feature of Fig.~\ref{fig:jt_secondb} is the strongly
anharmonic nature of the response in the absence of gravity: Fourier analysis
of the dashed red curve would clearly show significant power at many
frequencies. Given that the JT kernel $K(t,t')$ is obtained in the epicycle
approximation, in which all orbits are perfectly harmonic, and is driven by
a perfectly sinusoidal gravitational field, the anharmonic nature of the
response is remarkable. It arises through the integration of the sinusoidal
gravitational field along unperturbed orbits that carry stars in and out and
around in azimuth at a rate that speeds up or slows down as $x$ decreases or
increases. The self-gravitating response is less anharmonic, but the complex
structure of its minimum strongly suggests an interference pattern.

\begin{figure}
\centerline{\includegraphics[width=.9\hsize]{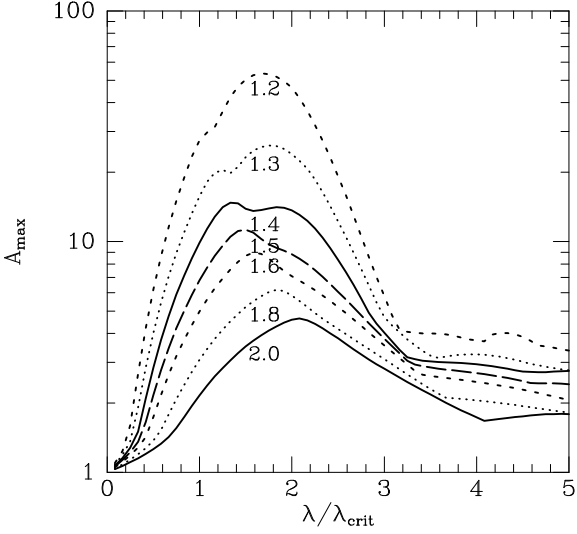}} \caption{The amount by
which the amplitude of the peak self-gravitating response $\Sigma_1(t)$
exceeds the peak response in the absence of self-gravity, $\Sigma_{\rm
s}K(t,t_\i)$ (eqn~\ref{eq:Amax}). Each curve is labelled by its value of $Q$. The individual data
points are obtained by searching for the value of the initial phase $\kappa
t_\i$ that maximises the plotted ratio.}
\label{fig:jt_secondc}
\end{figure}

\begin{figure*}
\includegraphics[width=.8\hsize]{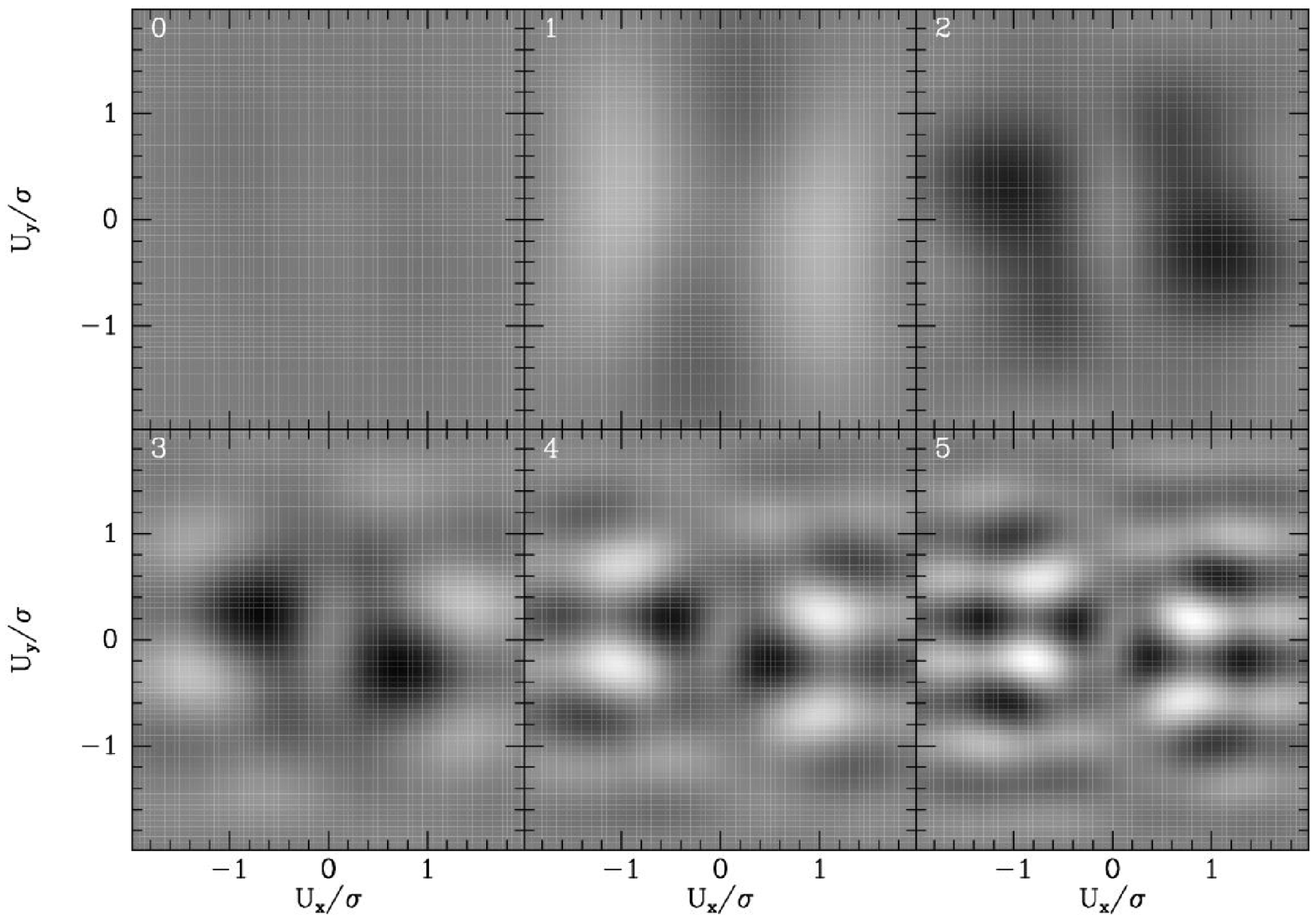}
\caption{The density of stars in velocity space at $(x,y)=(0,0)$ at six
times as a wave with $k_y=k_{\rm crit}$ swing amplifies in a sheet with
$Q=1.2$. The wave was initiated when $\kappa t=-\frac32\pi$, and from top
left to bottom right the panels show the perturbation to the 
velocity-space distribution of stars when   $\kappa
t/\pi=0,1,2,\ldots,5$. Equation (\ref{eq:foneSigma2}) was used to compute the
densities.}\label{fig:UV}
\end{figure*}

Fig.~\ref{fig:jt_secondc} plots for seven values of $Q$ the amplification
factor 
\[\label{eq:Amax}
A_{\rm max}\equiv\max_{t_\i}\biggl\{
{\max_t[\widetilde\Sigma_1(t)/\Sigma_{\rm s}]\over\max_t[K(t,t_\i)]}
\biggr\}
\]
 versus the lengthscale $\lambda/\lambda_{\rm crit}$. Here the outer max
operator involves a search over the time $t_\i$ at which the disc is jolted,
while the inner operators range over the times at which the magnitudes of the
responses are measured. The value of $t_\i$ that produces the maximum
response depends strongly on $\lambda/\lambda_{\rm crit}$. If the latter is
significantly less than unity, the disturbance phase mixes rapidly, so the
largest response arises when the disc is jolted shortly before $k_x$ passes
through zero, and $\kappa t_\i<\pi/2$. If $\lambda/\lambda_{\rm crit}\ga1$,
phase mixing is slower and the largest response is obtained by jolting the
disc early on, so there is time for a significant overdensity to emerge
before $k_x$ passes through zero. This is especially true if $Q-1$ is small.
Since maximising the response involves arranging for an oscillating system to
have a favourable phase at a particular instant, the optimum value of $t_\i$
is not a continuous function of $Q$ and $\lambda/\lambda_{\rm crit}$.
This fact gives rise to kinks and bumps in the curves of
Fig.~\ref{fig:jt_secondc} that might be thought indicatons of numerical error
but are not.

Fig.~\ref{fig:jt_secondc} closely resembles the central panel of Fig.~7 in
\cite{Toomre1981} although the quantity it plots is quite different. Toomre
(private communication) plotted the ratio of the rms responses obtained when
the disc was stimulated by leading and trailing waves with wavevectors that
differed only in the sign of $k_x$. The wavevector of the leading wave would
later become identical to the initial wavevector of the trailing wave, but
not before it had been swing amplified. Thus Toomre's ratio is a very clean
measure of the effectiveness of the amplifier.  By plotting the ratio of the
peak responses with and without self-gravity Fig.~\ref{fig:jt_secondc} is a
measure of the effectiveness of self-gravity.  The similarity of the two
figures indicates that the principal importance of self-gravity is in driving
the swing amplifier.

Fig.~\ref{fig:jt_secondc} shows
that at every value of $Q$, the amplification vanishes with the lengthscale
as a consequence of fast phase mixing, then rises to a peak
$\lambda/\lambda_{\rm crit}\simeq2$ before falling to a plateau that extends
from $\lambda/\lambda_{\rm crit}\simeq3$ to the longest lengthscales.
Discontinuities in the optimum value of $t_\i$ sometimes generate bumps in
the plateau. 

In N-body simulations of discs with sustained star formation, $Q$
settles to a value around $1.4$ \citep[e.g.][]{Aumer2016b}.
Fig.~\ref{fig:jt_secondc} indicates that in such a disc perturbations are
swing-amplified by a factor that hovers around 14 for
$1.3<\lambda/\lambda_{\rm crit}<2.1$.

Fig.~\ref{fig:UV} shows the evolution of the perturbation to velocity space
in a sheet with $Q=1.2$ when a wave that is stimulated by the surface density
(\ref{eq:impulseS}) at $t=-3\pi/2\kappa$ is
swing amplified. Successive panels show velocity space at intervals
$\pi/\kappa$ apart, starting with $t=0$, when $k_x=0$. The dotted curve in
Fig.~\ref{fig:jt_second} shows the evolution of this wave's density in real space.
This peaks before $t=3\pi/2\kappa$ and has long vanished by $t=4\pi/\kappa$. The
central panel in the bottom row of Fig.~\ref{fig:UV} shows that at this late
time the velocity-space signature of the wave is still growing. That is, the
wave persists for {\it much} longer in velocity space than it does in real
space.

In velocity space the wave manifests as a complex pattern of maxima and
minima that is constantly shrinking in scale velocity while increasing in
amplitude at a declining rate. The pattern varies with the location $\vx$ of
the plotted velocity space, becoming less symmetric as one moves away from
corotation. In general it looks like a network of cells.

\begin{figure}
\centerline{\includegraphics[width=.99\hsize]{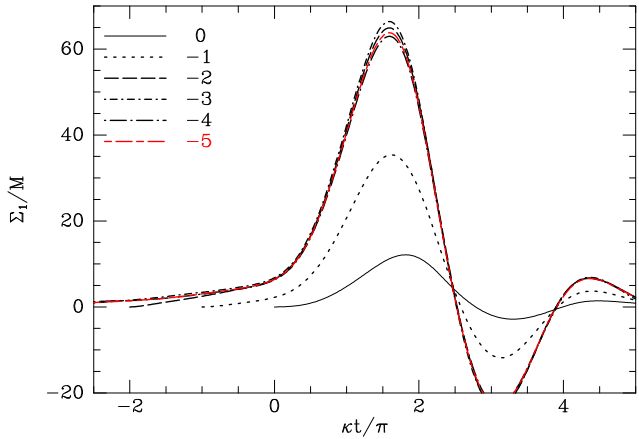}}
\caption{The temporal evolution of a wave with $k_y=k_{\rm crit}/2$ when a
mass $M$ is inserted at various times $t_\i$. The origin of time is the
instant at which $k_x=0$ and the legend gives the values of $\kappa t_\i/\pi$.
(cf.\ Fig.~5 of JT66).}\label{fig:jt1}
\end{figure}

\section{Response to a cloud}\label{sec:mass}

We now compute the response of a stellar disc to the insertion at time $t_\i$
of a mass such as a molecular cloud that moves on a circular orbit.  The mass
provides an endless succession of the $\delta$-function stimuli we considered
in the last section. 

Previously we considered stimuli that had a well defined circumferential
wavenumber $k_y$. Now we are considering a succession of broad-band stimuli
since when we Fourier transform the surface density
\[
\Sigma_{\rm e}(\vx)=M\,{\e^{-|\vx|^2/2\Delta^2}\over2\pi\Delta^2}
\]
we find that every wavenumber has non-vanishing amplitude:
\[
\widehat\Sigma_{\rm e}(\vk)\equiv\int\d^2\vx\,\e^{-\i\vk\cdot\vx}\,\Sigma_{\rm
e}(\vx) =M\,\e^{-|\vk|^2\Delta^2/2}.
\]
Notice that $\widehat\Sigma$ has dimensions of mass, unlike the quantity
$\widetilde\Sigma_1(t)$ defined above as the coefficient of
$\e^{\i\vk\cdot\vx}$ in the expression for $\Sigma_1(\vx,t)$. This difference
reflects the fact that to recover $\Sigma(\vx)$ from
$\widehat\Sigma(\vk)\e^{\i\vk\cdot\vx}$ we have to integrate over $\vk$.
Fortunately, for given $\vk$, $\widehat\Sigma_1$ will evolve in time in just
the same way that $\widetilde\Sigma_1$ does, so in the JT equation
(\ref{eq:Sigone}) we can replace $\widetilde\Sigma_1$ by $\widehat\Sigma_1$.
Then we have
\begin{align} \label{eq:SigM}
&\widehat\Sigma_1(\vk,t)
=\int_{t_\i}^t\d
t'\,K(t,t')\left[M\e^{-|\vk|^2\Delta^2/2}+\widehat\Sigma_1(\vk,t')\right].
\end{align}

Fig.~\ref{fig:jt1} shows solutions to equation (\ref{eq:SigM}) with
$\Delta=0$ for
$k_y=k_{\rm crit}/2$ and several values of the time $t_\i$ at which the
mass $M$ is added. The corresponding values of $\kappa t_\i/\pi$ are given in
the legend at top left. The quantity plotted is the amplitude of a swinging
wave as a function of time, with the
origin of time taken as the instant at which $k_x=0$, when the crests run
radially. The curve with the smallest amplitude is that for the latest time
of mass insertion $ t_\i=0$. The other curves show that as the moment of
insertion of the mass is pushed back, the amplitude of the wave grows until
it reaches peak amplitude for $\kappa t_\i/\pi\sim2.5$, and then settles to a
steady value for even earlier insertion. 

The mass $M$ will disturb the density in our coordinate patch with a
superposition of waves like that shown by the red curve in Fig.~\ref{fig:jt1}
that differ in the times at which $k_x=0$. If the moment of mass insertion
lies far in the past, every wave will be described by a curve that closely
resembles the red curve in Fig.~\ref{fig:jt1} but shifted to the right or
left according as its crests are vertical later or earlier than the wave for
which the red curve was computed. For this wave, $k_x=2Ak_yt$
(eqn.~\ref{eq:kx}). For a wave that has $k_x=0$ at time $t_x$, the relation
between $t$ and $k_x$ is different: $k_x=2Ak_y(t-t_x)$. Whereas in
Fig.~\ref{fig:jt1} there is one value of $k_x$ at each time, at any time
after insertion of the mass, the patch will contain waves with infinitely
many values of $k_x$ on account of the presence of waves that swing through
$k_x=0$ at every possible time. 

\begin{figure}
\centerline{\includegraphics[width=\hsize]{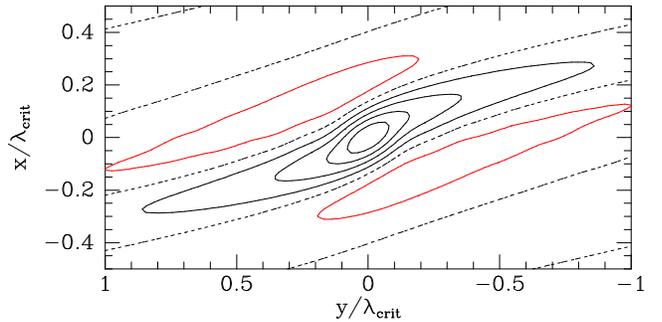}}
\caption{The overdensity created by a cloud with a Gaussian surface density
located at the origin in a disc with
$Q=1.4$. The contour values are $2,1.5,0.5,0$ and $-0.5$ times $M/(0.1\lambda_{\rm
crit})^2$. The zero contours are
dashed and the negative contours are red (cf.\ Fig.~7 of JT66).}\label{fig:wake}
\end{figure}

Whereas in the  section on impulsive excitation only one value of
$k_y$ was in play, the mass excites waves with every value of $k_y$.
The discussion of the last paragraph shows that from
Fig.~\ref{fig:jt1} we can infer the value of $\widehat\Sigma_1(\vk)$ along a
line of constant $k_y$ in the $(k_x,k_y)$ plane, and to determine
$\widehat\Sigma_1(\vk)$ over the the half plane $k_y>0$ we just need to solve
equation (\ref{eq:SigM}) for  positive values of $k_y$. Since
$\widehat\Sigma_1(\vk)$ is the Fourier transform of a real function, it
satisfies
\[\label{eq:star}
\widehat\Sigma_1(-\vk)=\widehat\Sigma_1^*(\vk),
\]
and this relation gives the values we need in the other half plane.

Fig.~\ref{fig:wake} shows contours of constant overdensity
\[
\Sigma_1(\vx)=\int{\d^2\vk\over(2\pi)^2}\,\e^{\i\vk\cdot\vx}\,\widehat\Sigma_1(\vk)
\]
for a cloud with characteristic size $\Delta=0.05\lambda_{\rm crit}$ in a
disc with a flat rotation curve and $Q=1.4$. The contour values are in units
of $M/(0.1\lambda_{\rm crit})^2$, so if the mass were spread at uniform
density it would contribute one unit within a square $0.1\lambda_{\rm crit}$
on a side. Since the innermost contour is at overdensity 2 in these units and
has an area that is roughly twice that of such a square, the excess stellar
mass within this contour alone amounts to four times the cloud's mass. From
this fact it is clear that a stellar disc like that of our Galaxy is very
polarisable: the effective mass of an object is several times its actual mass
on account of its tendency to cause passing stars to linger in its vicinity.

\section{Wave packets}\label{sec:packets}

\begin{figure}
\includegraphics[width=\hsize]{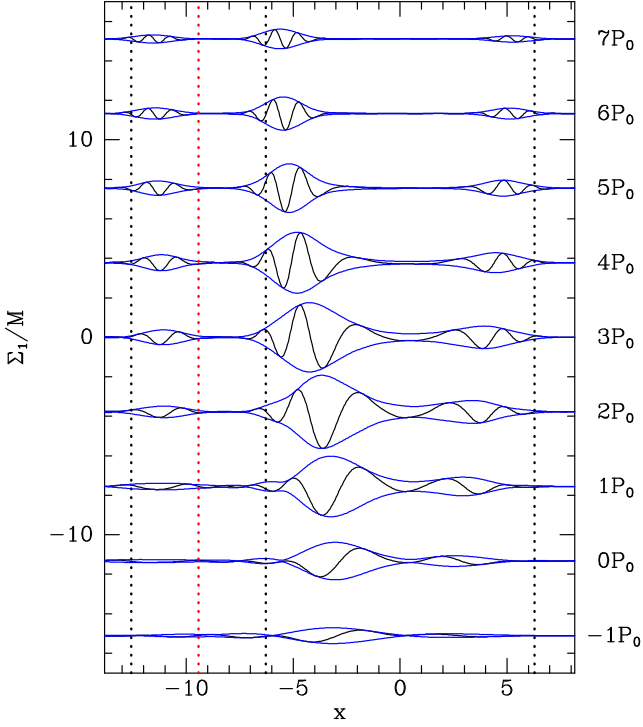}
 \caption{Propagation of a wavepacket through a sheet with $Q=1.2$. The
sheet's equilibrium is disturbed by an external density that has a Gaussian
radial profile and azimuthally varies as $\cos(k_\i y)$
(eqn.~\ref{eq:packet}) with $k_\i/k_{\rm crit}=\kappa/(4\pi A)=0.225$. From
right to left the black vertical lines mark the the outer and inner Lindblad
resonances at $\pm x_{\rm L}$ and the first harmonic of the inner Lindblad
resonance at $x=-2x_{\rm L}$, all for this value of $k_y$. The red vertical
line marks the centre of the structure at $x_\i=-1.5x_{\rm L}$. The black
curves show the coefficient of $\cos(k_\i y)$ in the response at nine times
separated by $P_0=2\pi/\kappa$ with the origin of time when the external
structure has peak density. The blue curves show the quadrature sums of the
coefficients of cos and $\sin(k_\i y)$. The units of length are those in
which $k_{\rm crit}=1$ (cf Fig.~3 of Toomre 1969).}\label{fig:packet}
\end{figure}

We now excite the disc with the potential of the mass
distribution
\[\label{eq:packet}
\Sigma_\e(\vx,t)={M\over\sqrt{2\pi}\Delta}\exp\left({(x-x_\i)^2\over2\Delta^2}\right)
\cos(k_\i y)\e^{-t^2/t_0^2}.
\]
The Fourier transform of the spatial part of this distribution is
\[\label{eq:FofM}
\widehat\Sigma_\e(\vk)=\pi
{M}\e^{-k_x^2\Delta^2/2-\i
k_xx_\i}[\delta(k_y+k_\i)+\delta(k_y-k_\i)].
\]
When this transform is used in the JT equation (\ref{eq:Sigone}) with
$\widetilde\Sigma$ replaced by $\widehat\Sigma$ as in the last section, both
$\widehat\Sigma_\e$ and $\widehat\Sigma_1$ become explicit functions of
$\vk$. Hitherto  the initial value of $k_x$ has been 
encoded via equation (\ref{eq:kx}) in the variable $t_\i$.   Now we
treat $t_\i$ as a function of $k_x(0)$:
\[
t_\i={k_x(0)\over2Ak_y}.
\]
As a consequence of this dependence of $t_\i$ on $\vk$, the upper limit of
the time integral in the JT equation loses its status as the current time. Instead we write
\[
t=t_\i+\tau,
\]
where $\tau$ is the current time. When evaluating
$\widehat\Sigma_\e(\vk,\tau')$ we use the current value of $k_x$:
\[\label{kx_tau}
k_x(\tau')=k_x(0)+2A\tau' k_y.
\]
When we inverse Fourier transform $\widehat\Sigma_1(\vk,\tau)$ to recover
$\Sigma_1(\vx,\tau)$ we likewise set $k_x$ to its value at time $\tau$.

\begin{figure}
\includegraphics[width=\hsize]{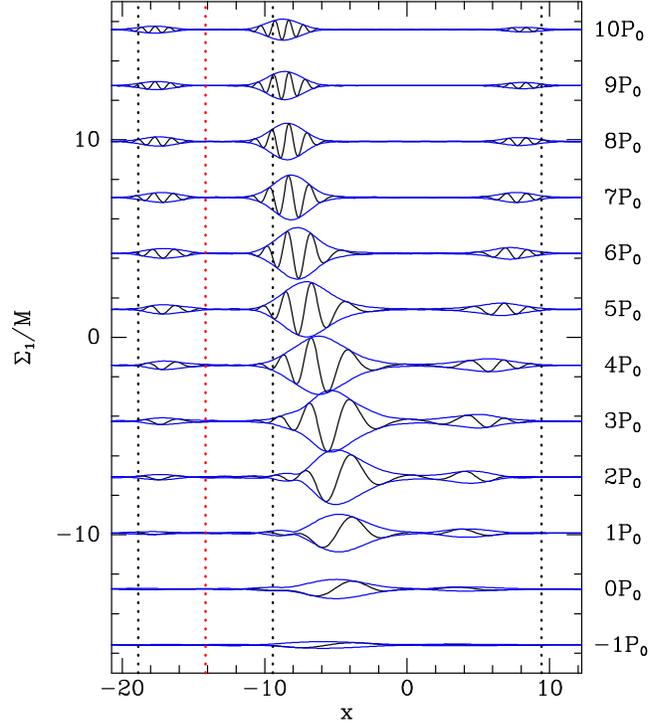}
\caption{The same as Fig.~\ref{fig:packet} except for $k_\i/k_{\rm
crit}=0.15$.}\label{fig:packet1}
\end{figure}

\begin{figure}
\includegraphics[width=\hsize]{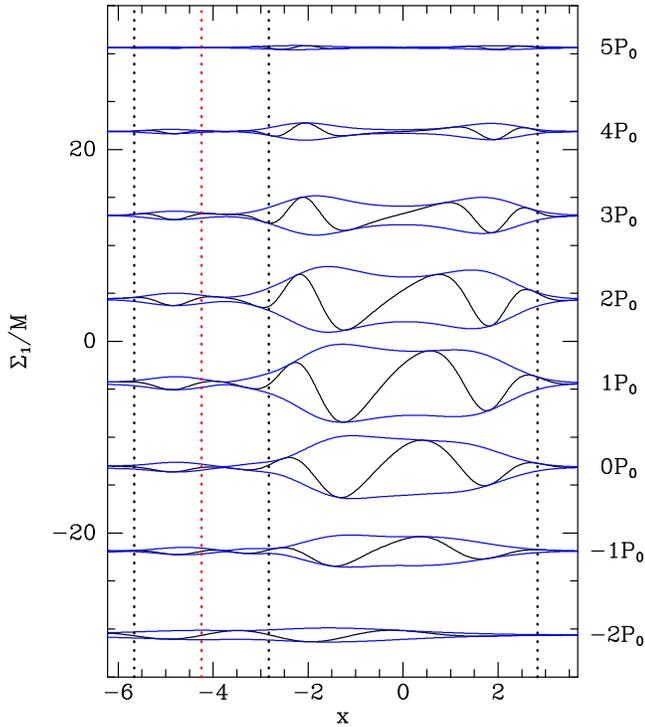}
\caption{As Fig.~\ref{fig:packet} except for $k_\i=k_{\rm crit}/2$. 
The vertical scale has been expanded by a factor $2$ to accommodate the
stronger response at this shorter wavelength. 
}\label{fig:packet2}
\end{figure}

In the previous applications of the JT equation, $\widehat\Sigma_\e$ has been
real, with the consequence that $\widehat\Sigma_1$ has remained real. Now
$\widehat\Sigma_\e$ is complex, so $\widehat\Sigma_1=\cR+\i\cI$ also becomes complex.
Using equation (\ref{eq:star}) we reason that
\begin{align}
\Sigma_1(x,t)&=\int_0^\infty\!{\d k_y\over2\pi}\int_{-\infty}^\infty\!\!{\d
k_x\over2\pi}\Bigl(\e^{\i\vk\cdot\vx}\widehat\Sigma_1(\vk,t)
+\e^{-\i\vk\cdot\vx}\widehat\Sigma_1(-\vk,t)
\Bigr)\cr
&=2\int_0^\infty{\d k_y\over2\pi}\int_{-\infty}^\infty{\d
k_x\over2\pi}\bigl[\cos(\vk\cdot\vx)\cR-\sin(\vk\cdot\vx)\cI)\bigr]\cr
&=\int_{-\infty}^\infty\!\!{\d
k_x\over2\pi}\Bigl\{\bigl[\cos(k_xx)\cR-\sin(k_xx)\cI\bigr]\cos(k_\i y)\cr
&\qquad-\bigl[\sin(k_xx)\cR+\cos(k_xx)\cI\bigr]\sin(k_\i y)\Bigr\},
\end{align}
where in the last line $\cR+\i\cI$ stands for the coefficient of
$\pi\delta(k_y-k_\i)$ in the transform $\widehat\Sigma_1(\vk)$
(cf.~eqn.~\ref{eq:FofM}).

Fig.~\ref{fig:packet} shows the result of exciting the disc with the external
density (\ref{eq:packet}) when $Q=1.2$, $\Delta=x_{\rm L}/\surd2$ and
$k_\i/k_{\rm crit}=\kappa/(4\pi A)=0.225$, where $x_{\rm L}$ is the distance
of the wave's Lindblad resonance from the corotation resonance
(eqn.~\ref{eq:defs_xL}). Although the exciting density is centred on
$x=-1.5x_{\rm L}$ (marked by a red vertical line), the excitation is
concentrated between the Lindblad resonances because the external density
corotates with particles at $x=0$, so stars that lie near the centre of the
exciting density pass it rather rapidly, and are less strongly perturbed than
more distant, but slower passing stars. 

The envelope of excitation first grows in amplitude and then bifurcates into
packets that propagate away from corotation inwards and outwards. The crests of
the waves are essentially stationary, but they wax and wane in such a way
that the packet moves quite coherently. That is, the phase velocity is much
smaller than the group velocity. The smallness of the phase velocity is a
direct consequence of the excitation being dominated by particles that nearly
corotate with the exciting density, which is stationary in our reference
frame.  As the packet moves and fades, $k_x$ gradually increases.
\cite{ToomreGp} showed that the trajectories of the packets' centres can be
accurately predicted from the group velocity implied by the LSK dispersion
relation.

If Fig.~\ref{fig:packet} a subsidiary packet develops {\it inside} the radius
$x_\i$ (marked in red) on which the exciting potential is centred. The left
edge of this packet lies near the value $x=-2x_{\rm L}$ at which stars on
circular orbits
perceive the wave at a frequency $\omega=2\kappa$. Our  explanation of this
feature relies on the structure of the axisymmetric limit of the JT kernel,
which is the subject of the next section.

\begin{figure*}
\includegraphics[width=.9\hsize]{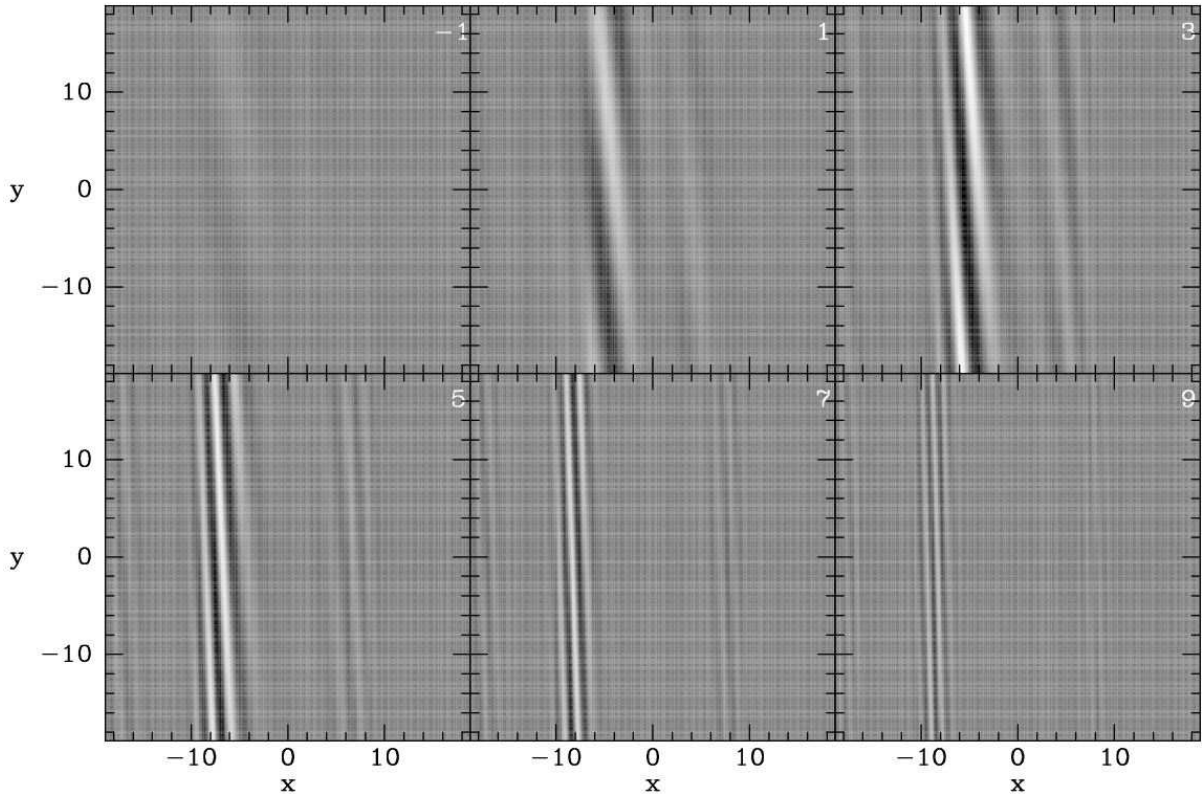}
\caption{The appearance in real space at six times (from top left to bottom
right $t=-1,1,3,5,7$ and $9P_0$)
of the wave packet with $k_y/k_{\rm crit}=0.15$ plotted in
Fig.~\ref{fig:packet1}. Black/white indicate $\Sigma_1/M=\pm1.6$.}\label{fig:wave_real}
\end{figure*}

Smaller values of $k_y/k_{\rm crit}$ than applies in Fig.~\ref{fig:packet}
yield similar wavepackets but ones that decay more slowly. The example shown
in Fig.~\ref{fig:packet1} has $k_y/k_{\rm crit}=0.15$. The largest packet
moves briskly to the inner Lindblad resonance but seems to get stuck with
about a quarter of its length over the resonance.  In this stationary
position, it slowly decays. Fig.~\ref{fig:wave_real} shows how this
disturbance looks in the $xy$ plane at six of the times plotted in
Fig.~\ref{fig:packet1}. The growth in the wave's amplitude up to
$t\sim3P_0=6\pi/\kappa$ is evident, as is the steady increase in $k_x$. One
can also see the tendency of both ingoing and outgoing packets to stick when
they reach a Lindblad resonance.

Fig.~\ref{fig:packet2} shows the effect of increasing $k_y$ to $k_{\rm
crit}/2$. At this shorter azimuthal wavelength, the disc is significantly
more responsive (Fig.~\ref{fig:jt_secondc}) and the disturbance stimulated by
the same mass is $\sim$twice as big. The stimulated spiral decays
significantly faster: the response has almost extinguished by $t=5P_0$ rather
than clearly persisting to $t=7P_0$ in Fig.~\ref{fig:packet} or to $t=10P_0$
in Fig.~\ref{fig:packet1}. The Lindblad resonances of this wave lie much
closer to corotation than do the Lindblad resonances of the previous waves
with lower $k_y$. In consequence the generated wavepacket fills the region
between the Lindblad resonances, and, perhaps because it touches both
Lindblad resonances from the outset, it shows little tendency to move.

\subsection{Application of the LSK dispersion relation}

In Section~\ref{sec:axisymm} we showed that the axisymmetric limit of the JT
kernel gives rise to a dispersion relation (\ref{eq:JTdisp}) that is identical to the
axisymmetric limit of the LSK dispersion relation.
We now use this  relation to elucidate the wavepackets
evident in Figs.~\ref{fig:packet} and \ref{fig:packet1}.

\begin{figure}
\includegraphics[width=\hsize]{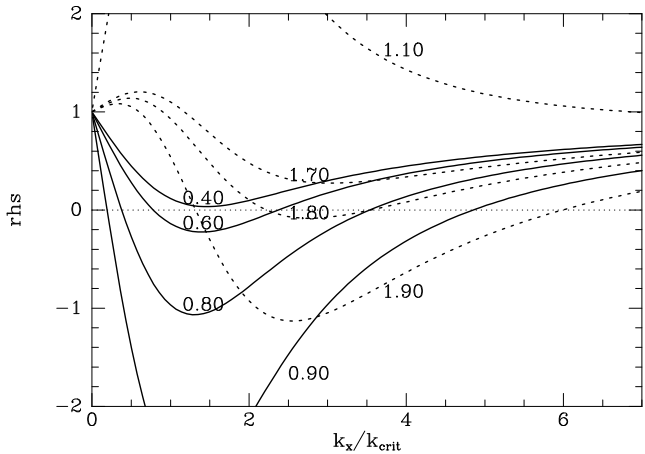}
\caption{The right-hand side of equation (\ref{eq:JTdisp}) plotted as a
function of $k_x$ for $Q=1.2$ and several values of $s$. Each curve is
labelled by its value of $s$.}\label{fig:JTdisp}
\end{figure}

Fig.~\ref{fig:JTdisp} plots the right-hand side of equation (\ref{eq:JTdisp})
as a function of $k_x$ for $Q=1.2$ and several values of $s$. The full curves
correspond to $s=0.4,0.6,0.8$ and $0.9$, while the dashed curves correspond
to $s=1.1,1.7,1.8$ and $1.9$. 
If the curve for some $s$ crosses the dashed line $y=0$, then there are
modes with that frequency at the two values of $k_x$ at which the curve
crosses $y=0$.
On account of the denominators $(n+1)^2-s^2$ in
equation (\ref{eq:JTdisp}), the curves change discontinuously as $s$
increases past an integer. Thus while the curve for $s=0.99$ (not plotted)
drops far below $y=0$ and thus yields modes at both small and large $k_x$,
the curve for $s=1.01$ (also not plotted) stays far above $y=0$ for all
$k_x$. Thus we have two bands of $s$ for which modes exist: $1>s>0.41$ and
$2>s>1.78$. Since the dispersion relation is a function of $s^2$, there are
equivalent bands with $s\to-s$.

We now argue that a disc will respond to tightly wound spirals very much as
it does to axisymmetric disturbance with the same value of $k_x$. On this
understanding, we now consider 
\[
s=m{\Omega_{\rm p}-\Omega\over\kappa} 
\]
to be a function of $x$, namely the frequency at which an $m$-armed spiral is
perceived by a star on a near-circular orbit.  The two bands in $s<0$ for
which modes exist thus correspond to two bands in $x$ within which
self-sustaining oscillations are possible. In Figs.~\ref{fig:packet} and
\ref{fig:packet1}, the band $-1<s<-0.41$ occupies territory to the right of
the middle vertical line, which marks the inner Lindblad resonance, while the
band $-2<s<-1.78$ occupies territory to the right of the left vertical line,
which marks the harmonic of the Lindblad resonance. The band $-1<s<-0.41$
should extend 60\% of the distance from the middle vertical line to $x=0$,
while the band $-2<s<-1.78$ should extend 56\% of the distance from the left
black vertical line to the red line. In Figs.~\ref{fig:packet} and
\ref{fig:packet1} the wavepackets  somewhat exceed these bounds, but qualitatively the
agreement is good. 

Hence we can interpret the wavepackets seen in
Figs.~\ref{fig:packet} and \ref{fig:packet1} as follows. The disturbance
created by the external driving potential $\Phi_\e$ is largest and lives longest
at locations where a self-sustaining mode is possible at the frequency at
which stars on nearly circular orbits perceive $\Phi_\e$. At $x<0$ these
locations are bounded on the left by $x=-2x_{\rm L}$ and $x=-x_{\rm L}$,
while at $x>1$ they are bounded on the right by $x=x_{\rm L}$. Since
$\Phi_\e$ has finite width in frequency space, modes with finite ranges in
$k_x$ are excited. Interference
between these modes first causes the region of largest net amplitude to
migrate towards the bounding lines $x=\pm x_{\rm L}$, $x=\pm2x_{\rm L}$, etc, and then causes
the net disturbance to fade as different modes drift more and more out of
phase and cancel ever more completely.

\section{Discussion}\label{sec:discuss}

A priori it is not obvious that in a warm, self-gravitating disc a
perturbation with the initial form $\Sigma_1(\vx)\propto\e^{\i\vk\cdot\vx}$
will evolve such that $\Sigma(\vx)$ remains proportional to
$\e^{\i\vk\cdot\vx}$ with $\vk$ evolving so the phase is constant at
particles that are on circular orbits. In our treatment this is established a
posteriori by showing that we {\it can} construct a solution to the
linearised CBE under the ansatz that $\vk$ evolves by simple shear. JT66
established this fact more cleanly by Fourier transforming the CBE and then
identifying its characteristics. 

The existence of the JT66 solutions to the linearised perturbation problem
clarifies the status of LSK running waves. The non-existence in a disc with
$Q>1$ of modes with values of $s^2$ smaller than a threshold value is
interpreted as an absence of waves in a region around corotation. As we have
seen, the JT equation establishes that axisymmetric modes exist at all radii.
The LSK dispersion relation has no solutions in a region around corotation
precisely because any non-axisymmetric wave must wind up: the wavevector must
be an explicit function of time, a possibility that one excludes a priori in
the derivation of a dispersion relation. In light of this remark, the
existence of solutions to the WKB dispersion relation away from corotation
becomes puzzling. The puzzle is resolved by the dynamics of wavepackets,
which \cite{ToomreGp} showed move towards the Lindblad resonances.
Consequently, a packet's central $\vk$ value becomes a function of time
through the dependence of the LSK dispersion relation on radius. In this way
the LSK theory manages after a fashion to encompass the growth in $k_x$ that
is a categorical imperative in a shearing system. 

Nevertheless, LSK waves do {\it not} play in discs a role analogous to
Maxwell's electromagnetic waves in electrodynamics.  Electromagnetic waves
are typically generated by shaking a charge, just as a solution of the JT
equation can be generated by gravitationally jolting the sheet. Once
generated, the waves of Section~\ref{sec:four} are free oscillations of the
system, exactly as a light wave is as it moves from emitting to absorbing
atom. The standing wave in a laser is generally imagined to comprise
counter-propagating travelling electromagnetic waves that are inter-converted
the the laser's end mirrors.  If spiral structure can be understood in terms
of normal modes, it will be by imagining the disc to be buzzing with JT66
winding waves, not LSK waves. The JT66 waves are simply as close as one can
get to normal modes in a shearing sheet. True modes don't exist because the
general JT kernel is not time-translation invariant.

In the special case of axisymmetry, waves don't wind up, so $k_x$ is
time-independent and the JT equation admits normal modes. Their dispersion
relation is identical with that of LSK waves with $m=0$.

We have explored the dynamics of wavepackets predicted by the JT equation by
using the method \cite{ToomreGp} devised to excite wavepackets comprising a
coherent group of waves that share a common value of $k_y$. When the
stimulating density has a small value of $k_y$ ($\la0.5k_{\rm crit}$),
packets form either side of corotation and move inwards and outwards to the
Lindblad resonances just as the LSK dispersion relation predicts.  If the
stimulating density has a larger value of $k_y$, the sheet responds more
energetically and in a less localised way. The stimulated structure decays
quite rapidly without significant motion away from corotation.  LSK theory
provides no insight into this behaviour.

A feature of wave patterns with reasonable small values of $k_y/k_{\rm crit}$
that \cite{ToomreGp} overlooked, is a wavepacket that fills the region at
$-2<x/x_{\rm L}<-1$, which lies inside the Lindblad resonance at $x=-x_{\rm
L}$. This newly identified wavepacket can be comprehended on the basis of the
LSK dispersion relation to the same extent as the wavepacket in $-1<x/x_{\rm
L}<0$ that \cite{ToomreGp} first displayed. The new wavepacket is a
nonetheless a curious phenomenon physically, because its surface density
oscillates faster than can any of the stars from which it is built. Moreover,
it is unlikely that any real galactic disc displays an excitation of this
type because in a real disc the inner Lindblad resonance is much further
removed from corotation than is the outer Lindblad resonance, and the analogue
of $x=-2x_{\rm L}$ would lie at an exceedingly small radius. The shearing
sheet, unlike a real disc, has resonances that are symmetrically arranged around corotation
because the epicycle frequencies $\kappa$ of its particles are independent of
$x$, whereas those of stars are roughly proportional to $1/R$.
 
Any wavepacket ultimately ceases to be visible in real space. We have seen,
however, that in velocity space it lives on. However, because first-order
perturbation theory does not encompass resonant trapping, it provides an
inadequate account of the manner in which waves decay.  N-body simulations
\citep{Se12} and calculations based on matrix mechanics
\citep{FouvryPMC2015} show that a wave is absorbed by stars that
resonate with the wave. Such stars are concentrated around the radii of the
Lindblad resonances. The concentration of the wave's energy on altering the
orbits of a small fraction of the disc's stars plays a key role in the
long-term drift of the disc to bar formation \citep{SellwoodC2014}, so it is
a pity that this physics is missed by the current theory. To capture this
phenomenon one needs recognise that the perturbation modifies a star's
frequencies in parallel with its orbit. When this fact is omitted, each
star's phase relative to the wave increments steadily, so the star spends a
pre-defined time at phases that cause it to absorb energy from the wave,
followed by an equal time during which its phase leads to energy being
surrendered. After a star becomes resonantly trapped, its phase relative to
the wave only librates, and normally is such that energy is absorbed from the
wave rather than being emitted.

Equation (\ref{eq:SigM}) for the perturbation caused by an orbiting mass
implies non-linear dependence on $\Sigma_0$ but linear dependence on $M$: we
can create a new solution by doubling both $M$ and $\widehat\Sigma_1$. It
follows that {\it any} object that moves on a circular orbit will have an
effective mass that is significantly greater than its actual mass; this
property is not restricted to massive objects such as GMCs.  It applies also
to stars so long as they move on circular orbits. While few stars do move on
perfectly circular orbits, radial oscillations with amplitude significantly
smaller than the extent of the wake shown in Fig.~\ref{fig:wake} should not
prevent the formation of a wake. Indeed, \cite{Toomre1991} showed that in
N-body simulations, regions of enhanced density like that shown in
Fig.~\ref{fig:wake} can be detected around individual stars if you stack
images of the sheet such that a different star always lies exactly at the
centre of the image. \cite{FouvryPMC2015} showed that the enhancement of the
masses of individual stars through wake formation accelerates the relaxation
of discs by increasing the level of Poisson noise. Much earlier,
\cite{Julian1967} pointed out that GMCs would stochastically accelerate
disc stars faster than one would naively expect because their effective
masses are larger than their physical masses by virtue of the wakes they
raise in the stellar disc.

Giant molecular clouds (GMCs) are thought to be destroyed by outflows from
the massive stars that they bring into the world. Hence it is interesting to
ask about the persistence of the wake that a GMC generates in the stellar
disc.  After all, a GMC gathers around it a stellar entourage significantly
more massive than itself. Will not this entourage seed an overdensity after
the GMC has been dispersed? Somewhat counter-intuitively, the current theory
predicts that the wake will quickly disperse rather than survive the GMC's
destruction. The argument is that, as JT66 demonstrated, each spiral wave
evolves in isolation, and the wake comprises waves that are already trailing.
For the wake to persist, a source is required of leading waves that can be
swing amplified.  So long as the GMC lives, its gravitational field furnishes
that source, but the wake cannot self seed.

The dynamical independence of waves with different $\vk$ values must derive
from the orthogonality of $\e^{\i\vk\cdot\vx}$ and $\e^{\i\vk'\cdot\vx}$ for
$\vk\ne\vk'$. From a physical perspective this orthogonality is questionable
because it hinges on an infinite domain of integration over $\vx$, which is
entirely unphysical.  Consequently, the implication of the JT equation that
GMC's after images are short-lived should be accepted cautiously.

The central principle of the linear theory used here is integration along
unperturbed trajectories. Such an integration will yield useful results so long
as the perturbation has not changed the orbit qualitatively. In the case of a
mass $M$, some orbits will be qualitatively changed by being trapped by the
corotation resonance. For the results computed to be useful, $M$ must be
small enough that only a small  fraction of stars are trapped by
the corotation resonance \citep[e.g.][]{Binney2018}.

\section{Conclusions}\label{sec:five}

Spiral structure is a complex phenomenon and decades have been required to
achieve a reasonable understanding of it. That understanding has been
assembled by patching together insights from several different approaches.
Matrix mechanics \citep{Kalnajs1977,Toomre1981,FouvryPMC2015} and N-body
simulations \citep[e.g.][]{SellwoodC2014,FouvryPMC2015} provide the most
trustworthy information, but taken alone they yield insufficient insight into
the physics of the phenomenon. Consequently, our understanding of the physics
of spiral structure is heavily dependent on linear theory \citep{Toomre1981}.
This comes in two variants and requires the use of approximations that are
not strictly justifiable. 

The LSK theory of running waves is quite well known but it requires the
tight-winding approximation, which inevitably fails as leading waves approach
corotation.  Moreover, taken on its own, the LSK theory is profoundly
misleading in that it draws an exclusion zone around corotation and suggests
that waves that approach this zone bounce off it.  Matrix mechanics and
N-body simulations clearly show that far from being a quiet zone, the
corotation region lies at the heart of spiral-structure.

The shearing sheet that JT66 introduced to stellar dynamics following the
seminal paper of \cite{GoldreichDLB} on gas discs, provides the only
tractable model of this beating heart of the system, so deserves to be widely
understood. It provides a tractable model of a spiral feature that includes
the vital time dependence of its wavevector $\vk$.  We have re-derived and
slightly extended the key results of JT66, and illustrated their value by both
reproducing important figures from JT66, \cite{ToomreGp}  and
\cite{Toomre1981} and by plotting
some things that cannot be found in those papers. Our aim has been twofold:
(i) to derive the JT equation in a way that is as self-contained and
elementary as possible, and then (ii) to explain more fully than JT66 and
\cite{ToomreGp} did how this equation can be used to compute a variety of
phenomena. These include,  the wake that an
orbiting body assembles around it, the movement of wavepackets between
corotation and Lindblad resonances, and the criterion for a disc's stability
to axisymmetric disturbances. All these applications are quite subtle
and merit explicit explanation.

Axisymmetric disturbances cannot wind up, so the wavevector $\vk$ is not
a function of time. Consequently, the axisymmetric limit
of the JT equation admits modes, and their dispersion relation is just the
appropriate LSK relation.

Application of the theory to wavepackets clarifies the standing of
the LSK dispersion relation. For sufficiently small azimuthal wavenumbers
($k/k_{\rm crit}\la0.5$), wavepackets move towards Lindblad resonances while
becoming more tightly wound, much as the LSK relation predicts. But for
larger wavenumbers, the LSK relation is of no use. The fundamental problem
with LSK theory is that in a shearing system, waves {\it will} wind up, so
$\vk$ should be an explicit function of time, yet such a functional dependence
is explicitly excluded in the derivation of a dispersion relation.

Everything here flows from the Volterra integral equation that JT66 first
derived, which admits inexpensive numerical solution. Only in the
axisymmetric limit $k_y\to0$ is the JT kernel $K(t,t')$ translationally
invariant, i.e., a function $\cK(t-t')$. From this it follows that in the
non-axisymmetric case a modal analysis such as that pursued by
\cite{LinShu1966} is impossible. The ingredients required to derive the JT
equation are (i) the linearised Boltzmann equation, (ii) the approximate
solution to Poisson's equation for short wavelength waves, and (iii) formulae
for general unperturbed orbits that are simple enough to permit analytic
integration over velocities to obtain the perturbed density from the
perturbed DF. Within the epicycle approximation, all of these ingredients
appear to be as available for a true disc as the shearing sheet. If it proves
impossible to generalise the JT kernel to a disc, the realism of the kernel
might still be significantly enhanced by making the epicycle frequency $\kappa$ a
decreasing function of $x$ and thus moving corotation towards the outer
Lindblad resonance   from the midpoint of
the gap between the Lindblad resonances. 

A striking result obtained here is the clarity with which a swing-amplified
wave can be seen in velocity space long after it has disappeared from real
space. The Gaia mission \citep{GaiaKatz2018} makes it straightforward to
probe in great detail the structure of velocity space at locations up to
$\sim2\kpc$ from the Sun. It has been known since the Hipparcos mission that
the local $UV$ plane is rich in structure \citep{WD98}, but universally
accepted explanations of this structure are still lacking. The shearing sheet
will surely play a prominent role in efforts to understand
velocity-space structures in the new data.

\section*{Acknowledgements}

I thank Jerry Sellwood and Scott Tremaine for comments on an early draft and
Alar Toomre for being a most helpful, meticulous and understanding referee.
This work  has been
supported by the UK Science and Technology Facilities Council under grant
number ST/N000919/1.

\bibliographystyle{mn2e} \bibliography{/u/tex/papers/mcmillan/torus/new_refs}

\appendix
\section{Gaussian integral}\label{app:A}
We have
\begin{align}
\int\d u\,(u+c)\e^{-a^2u^2+2\i bu}&=\e^{-b^2/a^2}\int\d u\,(u+c)\e^{-(au-\i
b/a)^2}\cr
&={\surd\pi\e^{-b^2/a^2}\over a^2}(ca+\i b/a).
\end{align}
Hence
\begin{align}\label{eq:app}
\int\d u_y\,&\e^{-a^2u_y^2+2\i b_yu_y}\int\d u_x\,\e^{-a^2u_x^2+2\i
b_xu_x}(c_xu_x+c_yu_y)\cr
&=
{\surd\pi\e^{-b_x^2/a^2}\over
a^2}\int\d u_y\,\e^{-a^2u_y^2+2\i b_yu_y}(c_yu_ya+\i c_xb_x/a)\cr
&={\i\pi\over
a^4}\e^{-(b_x^2+b_y^2)/a^2}( c_xb_x+ c_yb_y).
\end{align}

\section{Example code}\label{app:B}

The online version of this paper includes a file {\tt jt\_class.h}. Once this
header file has been included in a C++ program, the statements

{\obeylines\parindent=10pt\tt 
JulianT S(ky,Q); double A=S.Afactor(1000);
}

\noindent will place in {\tt A}, for a wave with $k_y={\tt ky*}k_{\rm crit}$
in a disc with stability parameter {\tt Q}, the amplification factor analogous to
those plotted in Fig.~\ref{fig:jt_secondc}. By default the disc has a
flat rotation curve, but this can be changed by the statement {\tt
S.set(Omega,kappa)} with the desired  frequencies. The current frequencies
are extracted by the statement {\tt S.fill(Omega,kappa,A)} 

The statements

{\obeylines\parindent=10pt\tt 
int np=1000;
double Sigma[np], theta[np];
S.Afactor(Sigma,theta,np);
}

\noindent will place in the arrays the quantities $\widetilde\Sigma_1(t)$
and $\kappa t$ shown by the black curves in Fig.~\ref{fig:jt_second}.
The statement

{\obeylines\parindent=10pt\tt 
S.Afactor(Sig,Sigma,theta,np);
}

\noindent will  additionally place in {\tt Sig[np]} data
for the red curve in Fig.~\ref{fig:jt_second}. All the above amplification
factors are for the case that the disc is jolted at time $t_\i$ such that
$\kappa t_\i=-3\pi/2$. The statement

{\obeylines\parindent=10pt\tt 
Afactor(Sig0,Sigma,theta,theta0,np);
}

\noindent will return the amplification factor, etc, for a disc jolted at
$\kappa t_\i={\tt theta0*}\pi$.

The statement {\tt S.K(theta,thetap)} will return the value of the JT kernel
at the given values of $\kappa t$, while {\tt S.Kt(t,tp)} will yield the
value of the kernel at the given times.

The statement {\tt S.Igrand(theta,thetap,Ux,Uy,dpsi)} returns the value of the
fraction on the second line of equation (\ref{eq:foneSigma2}) at the given values of
${\tt theta}=\kappa t$, ${\tt thetap}=\kappa t'$ and velocity components, and places
in {\tt double dpsi} the value of $\psi(t')-\psi(t)$. Hence the statements
 
{\obeylines\parindent=10pt\tt 
double Skr=sin(kx*x+ky*y),Ckr=cos(kx*x+ky*y);
double th=theta[850],dth=theta[1]-theta[0];
				double sum=-S.Igrand(th,0,Ux,Uy,dpsi);
				sum *=(Skr*cos(dpsi)+Ckr*sin(dpsi));
				for(int it=0;it<=nt;it++)\{
\quad					double thp=it*dth;
\quad					double I=S.Igrand(th,thp,Ux,Uy,dpsi);
\quad					sum-=Sigma[it]*I*(Skr*cos(dpsi)+Ckr*sin(dpsi))*dth;
				\}
				double f1=sum*exp(-.5*(Ux*Ux+Uy*Uy));
}

\noindent will place in {\tt f1} the perturbation to the density at {\tt(Ux,Uy)} in
the velocity space of {\tt
(x,y)} at the time corresponding to {\tt theta[850]}.

 \end{document}